\documentclass[fleqn,usenatbib]{mnras}

\usepackage{newtxtext,newtxmath}

\usepackage[T1]{fontenc}

\DeclareRobustCommand{\VAN}[3]{#2}
\let\VANthebibliography\thebibliography
\def\thebibliography{\DeclareRobustCommand{\VAN}[3]{##3}\VANthebibliography}

\usepackage{graphicx}	
\usepackage{amsmath}	

\usepackage{float} 
\usepackage{subcaption}
\usepackage{graphicx}

\title[ Rotationally inelastic rate coefficients for C$_7$N$^{-}$ and C$_{10}$H$^{-}$  in collision with H$_2$.]{ Rotationally inelastic rate coefficients for C$_7$N$^{-}$ and C$_{10}$H$^{-}$ anions in collision with H$_2$ at interstellar conditions}

\author[K. Giri et al.]{
K. Giri,$^{2}$
L. Gonz\'alez-S\'anchez,$^{3}$
F. A. Gianturco,$^{4}$\thanks{E-mail: Francesco.Gianturco@uibk.ac.at}
U. Lourderaj,$^{1}$ 
\newauthor
 A. Mart\'{i}n Santa Mar\'{i}a,$^{3}$
S. Rana,$^{1}$
N. Sathyamurthy,$^{5}$
E. Yurtsever,$^{6}$
and R. Wester$^{4}$
\\
$^{1}$School of Chemical Sciences, National
Institute of Science Education and Research (NISER) Bhubaneswar,\\
 An OCC of Homi Bhabha National Institute,  Khurda, Odisha 752050, India\\
$^{2}$Department of Computational Sciences, Central University of Punjab, Bathinda, Punjab 151401, India\\
$^{3}$Departamento de Química Física, University of Salamanca, Plaza de los Caídos sn, 37008 Salamanca, Spain \\
$^{4}$Institut für Ionenphysik und Angewandte Physik, Universität Innsbruck, A-6020 Innsbruck, Austria\\
$^{5}$Indian Institute of Science Education and Research Mohali, SAS Nagar, Punjab 140306, India\\
$^{6}$Department of Chemistry, Koc University Rumelifeneriyolu, TR 34450 Istanbul, Turkey
}

\date{Accepted XXX. Received YYY; in original form ZZZ}

\pubyear{2022}

\begin{document}
\label{firstpage}
\pagerange{\pageref{firstpage}--\pageref{lastpage}}
\maketitle

\begin{abstract}
The anions C$_7$N$^-$  and  C$_{10}$H$^-$ are  the two longest of the linear (C,N)-bearing and (C,H)-bearing chains which have so far been detected in the Interstellar Medium. In order to glean information on their collision-induced rotational state-changing processes, we analyse  the general features of  new ab initio  potentials describing the interaction of both linear anions with H$_2$, one of the most abundant partners in their ISM environment.  We employ  an artificial neural network fit  of the reduced-dimensionality potential energy surface for C$_7$N$^-$...H$_2$ interaction and discuss in detail the spatial features in terms of multipolar radial coefficients. For the C$_{10}$H$^-$...H$_2$ interaction we use the initial grid of two dimensional raw points to generate by quadrature the Legendre expansion directly, further including the long-range terms as discussed in the main text. Quantum scattering calculations are employed to obtain rotationally inelastic cross sections, for collision  energies in the range of 10$^{-4}$ to 400 cm$^{-1}$. From them we generate   the corresponding  inelastic rate coefficients  as a function of temperature covering the range from 10 to 50 K. The results  for the rate coefficients  for the longest cyanopolyyne are compared with the earlier results obtained for  the  smaller terms of the same series, also in collision with H$_2$. We obtain that the inelastic rate coefficients for the long linear anions  are all fairly large compared with the earlier systems. The consequences of such findings on their non-equilibrium rotational populations in interstellar environments are illustrated in our conclusions.
\end{abstract}




\section{Introduction}

In recent years, several linear (C,H)-bearing and (C,N)-bearing chains of  molecular anions have been detected at various sites in  interstellar media (ISM). Specifically: CN$^-$\citep{Agundez2010}, C$_3$N$^-$ \citep{Thaddeus2008}, C$_5$N$^-$ \citep{Cernicharo2008}, C$_4$H$^-$ \citep{Cernicharo2007, Agundez2008}, C$_6$H$^-$ \citep{McCarthy2006}, and C$_8$H$^-$ \citep{Brunken2007, Remijan2007}. The most recent observations of such linear anions have detected  two additional members  with the two longest sequences of carbon atoms  so far: C$_{10}$H$^-$  \citep{Rem2023} and C$_7$N$^-$ \citep{Cern2023}. They indeed  constitute  the most recent additions to the general class of molecules associated with the polyynes and cyanopolyynes anions. 

Astrophysical detection relies heavily on the spectroscopic investigations of such species in the laboratory and on matching sighted lines with those observed in the lab. To carry out  astrophysical modeling of the evolution of their distributions over a variety of internal states, important indicators are provided by the rate coefficients for the probability of rotational state-changes  induced by their collisions with He and H$_2$, both present in substantial amount in the same ISM environments. The collision-induced  occurrence of non-LTE (Local Thermal Equilibrium) population  for  rotational states in molecular partners could, in fact, be a significant path  to further radiative emissions from different excited spectral lines in the observational microwave regions.  Measurement of these rate coefficients in the laboratory is still challenging, while the available quantum mechanical methods can be used to compute them and use the results within  kinetic modelings of the underlying chemistry. 

Botschwina and Oswald \citep{Botschwina} had computationally determined the structural characteristics of some of these species to help observational sightings. For example, they found that the vibrational frequencies for C$_5$N$^-$  cover a broad range of values from 2235.5 cm$^{-1}$ down to 96.5 cm$^{-1}$.  Since the focus of our work is on rotational inelastic transitions in C$_7$N$^-$ and in C$_{10}$H$^-$  at low collision energies ($E_\text{coll}$ = 0.0001 to 400 cm$^{-1}$),  it can be safely assumed that their vibrational modes would hardly play any role at the energy regimes  considered in our present study. Furthermore, given the importance of the H$_2$ molecule as the most abundant molecular species in the same dark molecular cloud (DMC) environments, we shall focus our calculations on excitation and de-excitation processes in C$_7$N$^-$ and C$_{10}$H$^-$ induced by H$_2$, without including the state-changing processes for the hydrogen molecular partner. They will be considered, however, in a further, upcoming study.

  The chemistry of formation of such polyyne chains has also been the object of several studies and speculations (e.g. see: \citep{Cernicharo2020}. The possible formation of anions  from the neutral radical via a Radiative Electron Attachment (REA) process has been considered in some detail (as quoted in: \citep{18JeGiWe.Cnm} while a more direct chemical route by reaction of HC$_5$N with H$^-$ has also been put forward by our group \citep{15SaGiCa.LM} since our calculated rates were found to be large enough to be relevant within the chemical network producing these anions. In the final analysis, however, the various options indicated by the current literature have not yet coalesced into a unique proof of  the chemical formation of the fairly long  title anions, while however the calculated REA rates of reaction are still awaiting more sophisticated evaluations for such  linear chains.

 Rotational excitation and de-excitation in CN$^-$, the smallest term of the cyano-chain series, in collision with He and H$_2$ were investigated by Kłos and Lique \citep{Klos2011} and by some of the present authors \citep{GonzalezSanchezL}. Similar studies have been carried out  for the C$_3$N$^-$ $-$ He and C$_3$N$^-$ $-$ H$_2$ systems, by Lara-Moreno et al. \citep{Lara-Moreno2017, Lara-Moreno2019}. The next larger term in the series, the C$_5$N$^-$ in collision with H$_2$, has been discussed and analysed in our earlier work \citep{BGG23}.  On the other hand, there has been only very  recently a  study in our group \citep{LGS2023} that reported a quantum modeling for the collision-induced excitation/de-excitation of C$_7$N$^-$ rotational states by  He, while no  collision-induced rate coefficients have been computed for the longer term C$_{10}$H$^-$ of the series so far. These  rather long anionic systems provide a greater computational challenge, given the extended chain-lengths  and the much higher density of their rotational states which have to be considered in the quantum dynamics, as we shall further show below. As mentioned earlier, and considering  the expected ISM conditions in which such systems are found, we shall focus on rotational state-changing processes, thereby restricting this study to a rigid rotor treatment of both C$_7$N$^-$ and C$_{10}$H$^-$. 

 We should also bear in mind that the observational work on these two systems \citep{Rem2023},\citep{Cern2023} attributed six spectral lines arising from $\Delta$$j$= -1 transitions from rotational states $j$ = 27 through 32 from the dark cloud TMC-1 and 17 lines arising from transitions from 27 through $j$ = 43 to the molecular anion C$_7$N$^-$. While the former set of transitions could be identified using a rotational constant $B$ = 582.68490 +/- 0.00024 MHz and distortion constant $D$ = 4.01+/- 0.13 Hz, the latter set could be interpreted in terms of $B$ = 582.6827 +/- 0.00085 MHz and $D$ = 3.31 +/- 0.31 Hz. These constants matched nicely with the ab initio computed values of $B$ = 582.75 MHz and $D$ = 4.3 Hz for C$_7$N$^-$. The same quantum chemistry calculations yielded a value of 7.5 $D$ for the dipole moment of the anion. 
Remijan  et al.\citep{Rem2023} further analysed the spectral data from TMC-1 based on $B$ = 299.882 MHz obtained from density functional theoretic calculations using the M06-2X functional and the 6-31+G(d) basis set and scaled on the basis of the $B$ values obtained for the smaller anions in the series C$_{2n}$H$^-$. They attributed the spectral features to a $B$ value of 299.87133 MHz for C$_{10}$H$^-$, by considering $\Delta$$j$ =-1 transitions originating from $j$ = 14 through 40 in their spectral simulation.  The rotational temperatures in TMC-1 were estimated to be around 6-7K, while in the case of the IRC+10216 it was estimated to be around 26-29K.
Therefore, we have undertaken in the present study an ab initio dynamical investigation of the corresponding transitions in C$_7$N$^-$ and C$_{10}$H$^-$, as will be discussed in the later sections below, involving specifically some of the rotational transitions detected during the observational studies mentioned above. 

  The computed potential energy surfaces (PESs) of the present study can be considered either within an averaged-orientation scheme (see below) which then yields  a  two-dimensional (2D) PES,  or  as a full-orientation within a four-dimensional (4D) PES for the title anions  interacting with the H$_2$ partner. In the following Section we shall show in detail the formulation involving an orientation averaging of the H$_2$ molecular partner, while we shall describe and discuss both full 4D interactions in a future paper in preparation. Earlier studies on similar systems
  \citep{Giri} have shown that the computed inelastic rate coefficients  obtained within a 2D description of the relevant PES are usually within a few  percent of the exact evaluations in 4D.
\section{Computational Methods}

\subsection{The Computed Potential Energy Surfaces}
The geometry of the C$_7$N$^-$ anion was optimized using the  coupled cluster singles, doubles and perturbative triples (CCSD(T)) method and the cc-pVQZ basis set via the MOLPRO suite of computer programs \citep{molpro}. Its ground state was found to be a linear chain with the optimized bond distances already reported in our recent work  \citep{LGS2023}.  These computed features are in good agreement with those from the previous study by the Botschwina's group \citep{Botschwina}. The C$_7$N$^-$$-$H$_2$ interaction potential was computed  using the basis set described as the aug-cc-pVTZ for all the C, H and N atoms within the CCSD(T)-F12b framework. In many of our earlier studies on these anionic chains we have compared the -VQZ level of calculations with additional calculations carried out at the Complete Basis Set (CBS) extrapolation and found only small differences of about 1-2$\%$ in the regions around the minima of the attractive wells of the complex.

\begin{figure}
    \centering
    \includegraphics[width=0.45\textwidth]{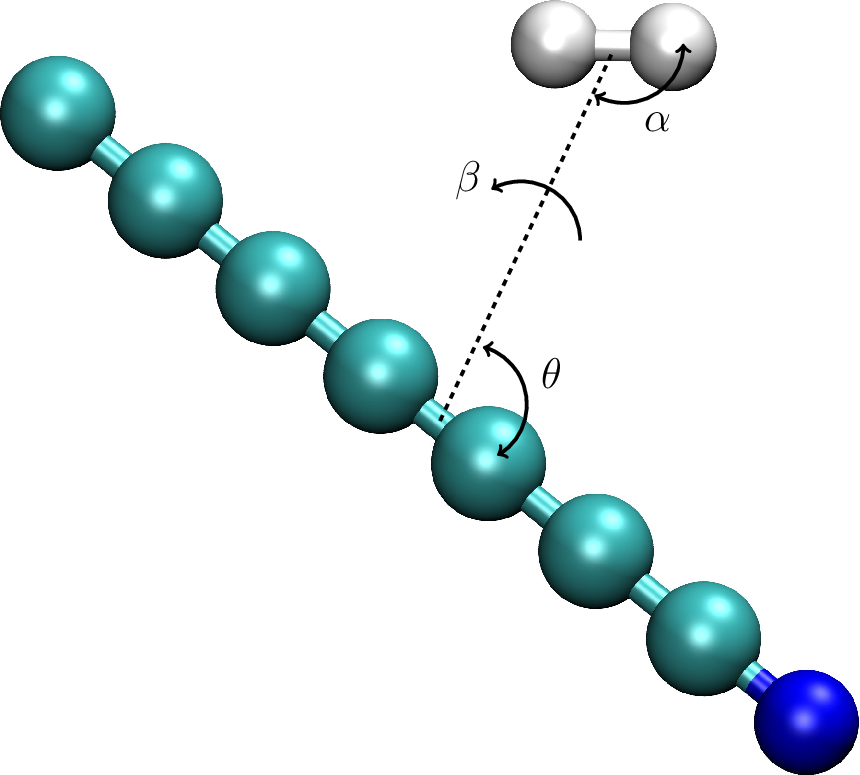}
    \caption{Coordinates  used to represent the C$_7$N$^-$H$_2$ system.The radial coordinate R is along the molecular length. The green spheres represent the carbon atoms, the blue sphere  the nitrogen atom and the grey spheres represent the hydrogen atoms. The same coordinates were also used in the case of the C$_{10}$H$^-$ anionic partner of the H$_2$ molecule, but not repeated in the present Figure.}
    \label{fig:figure1}
\end{figure}

The Jacobi coordinate system used to represent the C$_7$N$^-$$-$H$_2$ system is shown in Figure \ref{fig:figure1}. Here, $R$ is the distance between the center-of-mass of C$_7$N$^-$ (COM-pcy) and the center-of-mass of H$_2$ (COM-H$_2$), $\theta$ is the angle between COM-H$_2-$COM-pcy$-$N, $\alpha$ is the angle between COM-pcy$-$COM-H$_2-$H, and $\beta$ is the torsional angle about $R$.  The interaction energies between C$_7$N$^-$ and H$_2$ including the basis set superposition error (BSSE) correction were calculated at the CCSD(T)-F12b/aug-cc-pVTZ level of theory using MOLPRO \citep{molpro} by varying $R$ and $\theta$  and considering three orientations of H$_2$.  The  C$_7$N$^-$ and H$_2$ were treated as rigid molecules. The orientations considered were: (i) $\alpha$ = 90$^\text{o}$, $\beta$ = 0 (the 'perp' orientation), (ii) $\alpha$ = 0, $\beta$ = 0 (the 'in-line' orientation), and (iii) $\alpha$ = 90$^\text{o}$, $\beta$ = 90$^\text{o}$ (the out-of-plane, 'oop' orientation). 

 A  similar procedure has  been followed for the longer polyyne chain of the C$_{10}$H$^-$ anion, also interacting with the H$_2$ neutral molecule. More specifically, the longer anion's individual bond lengths were optimized in the isolated molecule using the methodology mentioned earlier, leading to an overall chain length of 12.636 \AA .  Since C$_{10}$H$^-$ is that long,  it is expected that  the intermolecular repulsion (with H$_2$) would be fairly high around $R$ = 6.5 \AA. In reality, the repulsion becomes high enough for $R$ < 8 \AA\ for certain orientations in the energy range investigated in the present study. Therefore, given this occurrence of highly repulsive features at such large distances, the potential energy had to be computed only for certain angles in the case of the shorter $R$ values as we were only interested in low-energy collision events, discussed more in detail in the following Sections. 
 
 As discussed for the previous C$_7$N$^-$ chain in this work, the interaction energies in the case of the longer anion also  included the BSSE correction and were calculated at the CCSD(T)-F12b/aug-cc-pVTZ level of theory using MOLPRO \citep{molpro}. The range of angles was   between 0 and 180$^\text{o}$ with increments of 5$^\text{o}$, while the $R$-grid was taken  from 8.0 \AA~ to 12.0 \AA~  with 0.1 \AA~ increments and then over the range of 12.0-15.0 \AA~  with 0.2 \AA~ increments. The total number of raw points was in this case about 4,150 within the short-range region of the interaction.

 To investigate the effect of averaging over different sets of orientations, we compared the average potential obtained using three orientations reported in the manuscript with that of the average over an overall total of nine orientations. The six additional orientations ($\alpha$,$\beta$) are: (45, 0), (45, 45), (45, 90), (90, 45), (135, 0), and (135, 45) in degrees. The differences in their energies for different values of $\theta$ were analysed in detail and we had found that the mean difference is  around 5.8 cm$^{-1}$. Slightly larger differences were seen only for $\theta$ = 180°, although that geometry will contribute little to the efficiency of the orientational dynamics of rotational energy transfer, as further discussed below. 
 
 The full 4D interactions were  therefore averaged to yield  two 2D, reduced-dimensionality PESs as discussed in detail in the next Subsection. The latter were then employed for the calculations discussed and reported in the present work, while the use of the 4D PESs will be reported later on in a different publication. Our own experience with the dynamics results via the dimension-reduction scheme, however, has been that the 2D-originated cross sections turn out to be only  within a few \% of the 4D-dynamics results.

\subsubsection{ The reduced-dimensionality PES for the C$_7$N$^-$$-$H$_2$ case.}

We generated a two-dimensional PES, $V(R, \theta)$,  by averaging over the three orientations as already performed in our earlier work  \citep{BM2021}:
\begin{equation}
V_\text{av}(R,\theta) = \frac{1}{3}( V_\text{perp}(R,\theta) + V_\text{in-line}(R,\theta)  + V_\text{oop}(R,\theta) ).
\label{eq.V_av}
\end{equation}

This treatment of the interaction with the H$_2$ partner is physically equivalent to restricting the latter molecule to be only the para-hydrogen in its $j$ = 0 level and has been already used in the past for various other systems, as we discussed in our earlier work \citep{BM2021}. For the temperatures of interest in the present study (T $\le$ 50 K) the probability of exciting the H$_2$ molecule is rather negligible as the energy spacing between the $j$ = 0 and $j$ = 2 levels of para-hydrogen is around 500 K, hence much larger than the energy gaps in the two long linear chains of the present anions.

\begin{figure}
    \centering
    \includegraphics[width=0.5\textwidth]{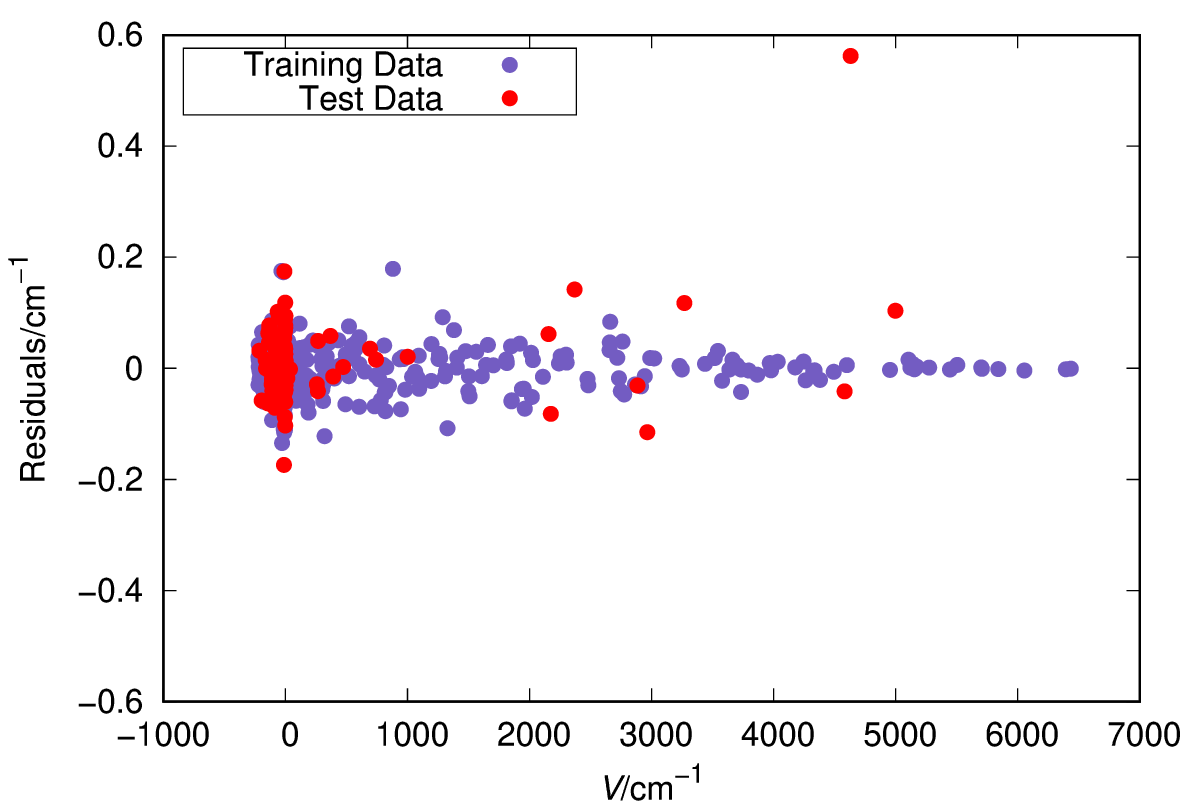}
    \caption{Residuals for the training and test data from the best ANN fit (Fit 4) of the 2D-PES for C$_7$N$^-$$-$H$_2$.}
    \label{fig:figure2}
\end{figure}

\begin{figure}
    \centering
    \includegraphics[width=0.5\textwidth]{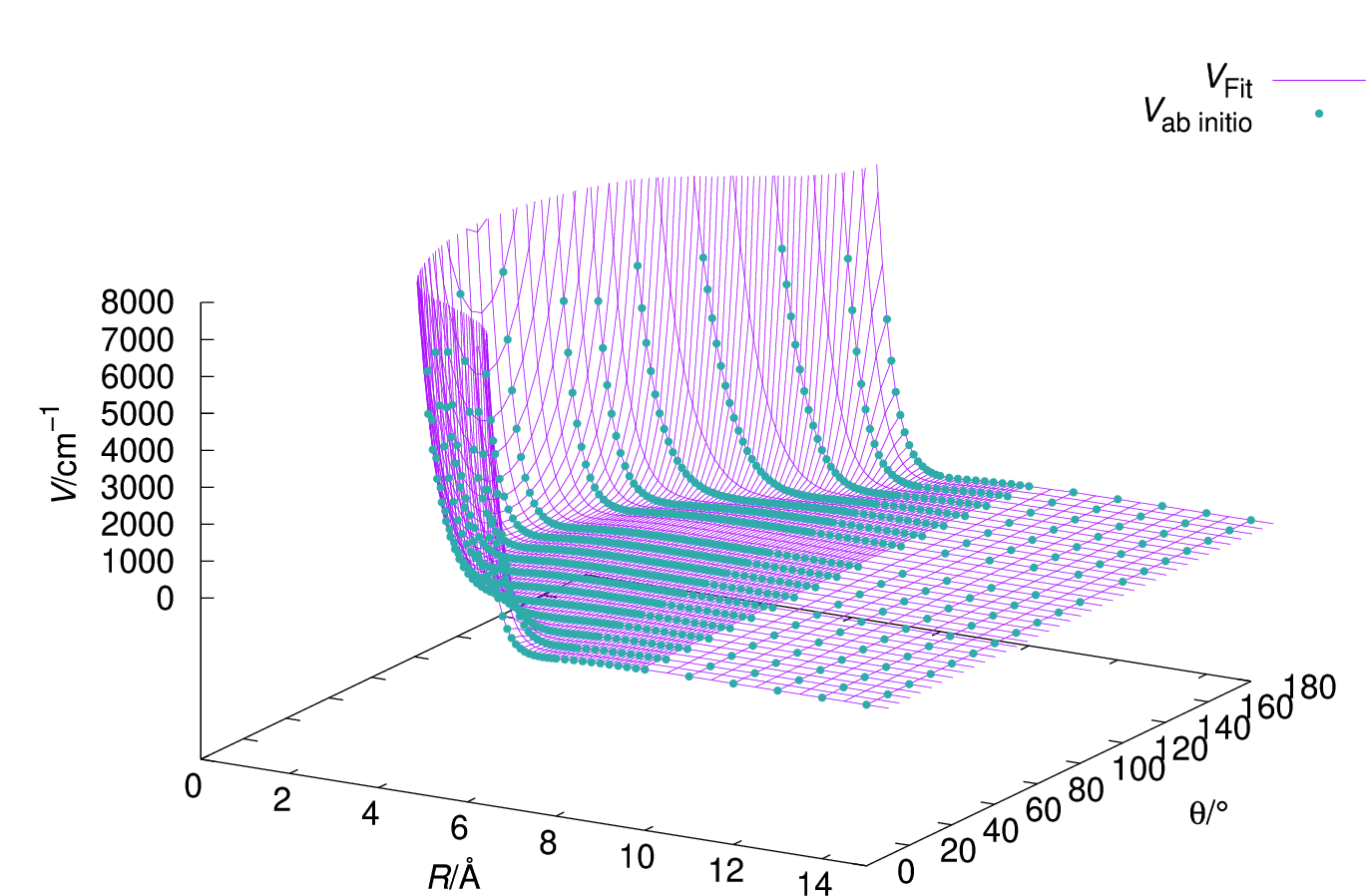}
    \caption{Plot of the PES in polar coordinates, obtained from the best ANN fit (Fit 4) compared with the ab initio data for the C$_7$N$^-$$-$H$_2$ system.The ab initio points are given by blue dots while the ANN fit is shown by the purple curves.}
    \label{fig:figure3}
\end{figure}

 For the C$_7$N$^-$ anion, the reduced-dimension ab initio data consisted of a total of 1,124 points. They were then fitted using the Artificial Neural Network (ANN) method \citep{raff2012, Sarkar, Biswas, Giri}. The data was divided into training and test sets in the ratio of 90:10 and several fits were obtained by considering different network sizes. The Bayesian regularization algorithm was used for  training and a target performance ($\chi^2$) was set at 0.001 cm$^{-2}$ for a maximum of 10000 epochs. The results of the various fits are summarized in Table \ref{tbl:2D_ANN_Fit}. The fits were found to be good with small RMSDs and the best fit (Fit 4) was obtained for a network size of (25, 25) with  RMSDs of 0.032 and 0.076 cm$^{-1}$, respectively, for the training and test data sets having a maximum deviation of 0.562 cm$^{-1}$ for the interaction energy of 4630 cm$^{-1}$ corresponding to $R =$ 2.1 \AA~ and $\theta$ = 100$^\text{o}$.  The residuals corresponding to Fit 4 are given in Figure \ref{fig:figure2}. It can be seen that overall, the deviations are within 0.2 cm$^{-1}$ for both the training and test data.  In addition, the ANN fit represented the PES smoothly over the entire range of $R$ and $\theta$ without showing any signs of overfitting, as illustrated in Figure \ref{fig:figure3}.  The best ANN fit (Fit 4) of the PES  was then used for the quantum dynamics simulations. 
 
\begin{table}
    \centering
    \caption{ANN Fits of the 2D-PES for C$_7$N$^-$$-$H$_2$}
    \label{tbl:2D_ANN_Fit}
    \begin{tabular}{cccccc}
    & & & & &\\
    \hline
    {ANN} & {Network} & \multicolumn{2}{c}{RMSD (cm$^{-1}$)} & {Maximum} & {No. of} \\
    \cline{3-4} 
    Fits  &    size   &   training     &    test           & Deviation (cm$^{-1}$) & Epochs \\ 
    \hline
    Fit 1 & 10, 10 & 0.089 & 0.158 & 0.855  &  10000\\
    Fit 2 & 15, 15 & 0.043 & 0.124 & 0.428  &  10000\\
    Fit 3 & 20, 20 & 0.035 & 0.086 & -0.565 &  10000\\
    Fit 4 & 25, 25 & 0.032 & 0.076 & 0.562  &  9860\\
    \hline
\end{tabular}
\end{table}

\begin{figure}
\centering
\includegraphics[width=1.0\linewidth]{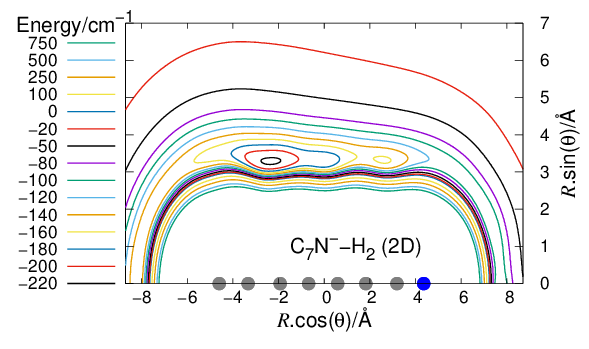} 
\caption{Potential energy contours for the C$_7$N$^--$H$_2$ system within the orientation-averaged 2D form discussed in the main text. The latter is given by the averaged combinations of the $\alpha$ and $\beta$ angles following the procedure already described in our earlier work \citep{BM2021}. The N-end of the anion is represented by a blue dot along the x-axis.}
\label{fig:fig4}
\end{figure}

The anisotropy of the potential for the C$_7$N$^-$$-$H$_2$ system is shown as a function of $R$ and $\theta$ by the  isolines of Figure \ref{fig:fig4} . That orientation angle has been already defined in the pictorial view in Figure 1. We also see from the data in Fig.4 that the deepest attractive well region is for the T-shaped geometry centered on the mid C-atom of the chain. This is what would be expected from the closed-shell nature of the neutral H$_2$, treated as a point mass object, that tends to avoid the negative charge at the N-end of the anion. Hence,our calculations found that the potential has a minimum of about $-$223.37 cm$^{-1}$ at $R$ = 4.0 \AA \, and $\theta=125.4^\text{o}$ indicating that near the T-shaped geometry we find the global minimum configuration for this system. In generating the initial raw points, the  $\theta$ steps were of 10$^\text{o}$, while the values of the radial parameter $R$, depending on the angular value, were varied from 6.4 \AA\ to 10.0 \AA\ with increments of 0.1 \AA.  Additionally,  we used  more distant radial  values from 10 \AA\ to  20 \AA\  at all angles with steps of 1.0 \AA.

 The final PES was expanded in orthogonal Legendre polynomials for the angle $\theta$ as the system is now described via two variables only: 

\begin{equation}
        V(R,\theta) = \sum{V_{\lambda}(R) P_{\lambda}(\cos\theta)  }
\label{eq:vl}
\end{equation}

\begin{figure}
\centering
 \includegraphics[width=0.7\linewidth,angle=-90]{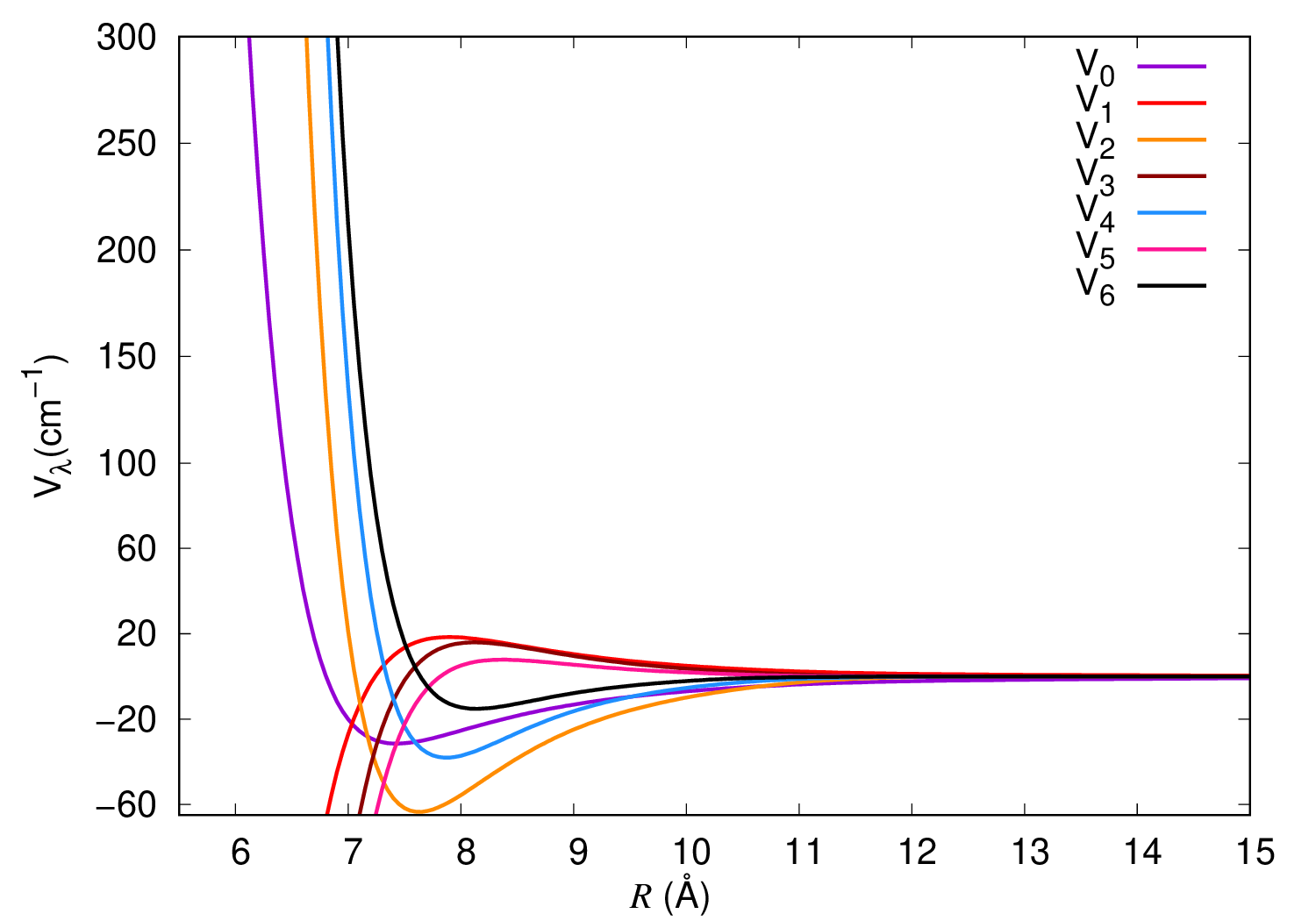} 
\caption{Plot of the Legendre coefficients as a function of $R$ for the   C$_7$N$^-$$-$H$_2$ system in the 2D reduced dimensionality. The H$_2$ molecule is treated as a structureless partner. The 2D-PES is derived from the  average of  combinations of $\alpha$ and $\beta$ angles as discussed in the main text.} 
\label{fig5}
\end{figure}

The number of ${\lambda}$ terms required for numerical convergence was extended up to ${\lambda}$ = 80, although at many of the collision energies sampled in our calculations only the lowest ten  terms were sufficient.  The strong angular anisotropy of the interaction potential is  confirmed when we examine the sample set of  radial  expansion coefficients  reported in Figure \ref{fig5}. We see that such coefficients exhibit markedly different radial regions of coupling strength which  depend on their ${\lambda}$ label, and which  strongly affect the relative efficiency of  state-to-state transitions as we shall further discuss in the following Section. This is because the ${\lambda}$ value controls the direct coupling between rotational quantum states within the quantum dynamics (see below), hence can tell us about the different excitation probabilities  for the involved state-changing processes. One has to bear in mind, however, that given the very small energy gaps between the transitions of interest here, in many cases such preferential rules will be acting less noticeably than in the earlier, shorter chain systems discussed by us.

      In any event, by examining the different radial terms reported in Figure \ref{fig5} we note that the ${\lambda}$ = 2 coefficient exhibits the largest attractive well region, indicating  a more effective, direct dynamical coupling between rotational states with $\Delta$$j_1$ = 2 in comparison with $\Delta$$j_1$ = 1 or = 3, as we shall further verify later with the actual calculated cross sections. One should keep in mind, however, that the exceptional smallness of the energy gaps between levels in both systems will show up with only small effects from the propensity rule above, as discussed again  when presenting our results.
      
   The asymptotically correct long-range potential $V_{\textnormal{LR}}$ for a dipolar species with dipole moment $\mu$ interacting with a closed shell atom/molecule with the polarizability $\alpha$ ($\alpha_{\parallel}$ and $\alpha_{\perp}$ being the parallel and perpendicular components) is given by, e.g.: \citep{LGS2023}.

\begin{equation}
V_{\textnormal{LR}} = - \frac{\alpha_0}{2R^4} + (\frac{2\alpha_0\mu}{R^5})\textnormal{cos}\theta, 
\label{eq:lr}
\end{equation}
where $\alpha_0$ = ($\alpha_{\parallel}$ + 2$\alpha_{\perp}$)/3.

Polarizability for H$_2$ at $r$ = 1.4 $a_{0}$ (0.7408 \AA) \citep{kol1967} is given by $\alpha_{0}$ = 5.1786 $a_{0}^3$ and $\alpha_2$ = 1.8028 $a_{0}^3$, although in the present extrapolation procedure, where the 2D reduction omits to treat the non-sphericity of the H$_2$ partner explicitly, we did not include this anisotropic term in the expansion. For comparison, $\alpha_{0}$ = 1.41 $a_{0}^3$ for He  (see: \citep{LGS2023}. For C$_7$N$^-$, $\mu$ = 7.545 D (2.953 $a_{0}^3$) (ref: \citep{Cern2023}.    

The final form of the potential ($V_{\textnormal{f}}$) using the switching function is given by,

\begin{equation}
    V_f = f_s V_{\textrm{ANN}}+(1-f_s)V_{\textrm{LR}}
\label{eq:vf}
\end{equation}
where the switching function is,

\begin{equation}
    f_s(R) = \frac{1}{e^{\frac{(R-R_0)}{\Delta R}+1}}
\label{eq:sf}
\end{equation}
and $R_0 = 11 $ \AA, and $\Delta R$ = 0.5  \AA. 

The R$_0$ and $\Delta$R parameters used for including the long-range (LR) potential terms have been optimized by trial and error based on previous cases we had studied. For every trial we check the smoothness of the fitted PES in the range where the long-range interaction becames relevant. Hence, the switching parameters R$_0$ and $\Delta$R were chosen by plotting the tail-end of the ANN potential and the asymptotically correct long-range potential and smoothly connecting the two in the switch-over region .

 
 \subsubsection{ The 2D PES for the  C$_{10}$H$^-$$-$H$_2$ System.} 

  For a 2D description of the PES for the longer polyynic anion interacting with H$_2$, we have employed the very same averaging procedure already outlined in the previous subsection, using about 4,000 raw points from the assembly of raw points discussed earlier in this work.
  
\begin{figure}
\centering
\includegraphics[width=0.9\linewidth,angle=+0.0]{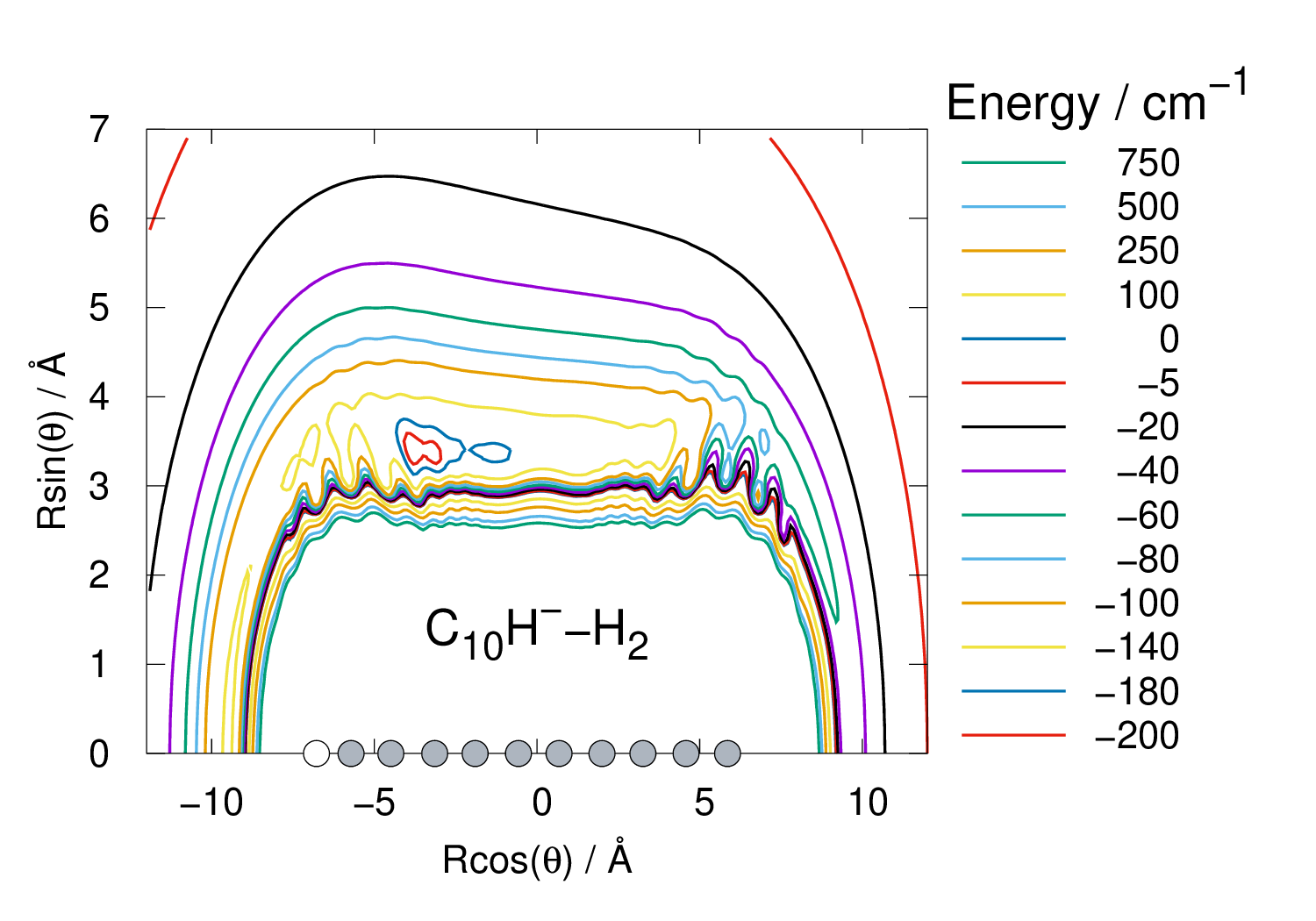} 
\caption{Potential energy contours for  the C$_{10}$H$^-$ $-$ H$_2$   system within the orientation-averaged 2D form discussed in the main text. It is obtained by the  averaged  combinations of the $\alpha$ and $\beta$ angles following the procedure described in our earlier work  \citep{BM2021}. The H-end of the anion is represented by a blue dot along the x-axis. See main text for further details.}
\label{fig6}
\end{figure}

  The resulting 2D-PES is plotted in Figure \ref{fig6} and it is seen that the minimum in the PES corresponds to a T-shaped geometry of the C$_{10}$H$^-$$-$H$_2$ complex. More specifically, the potential has a minimum of about $-$210.37 cm$^{-1}$ at $R$ = 5.0 \AA \, and $\theta=138.0^\text{o}$ indicating that near the T-shaped geometry we find the global minimum configuration, as we did in the previous case.   

\begin{figure}
\centering
 \includegraphics[width=0.9\linewidth,angle=+0.0]{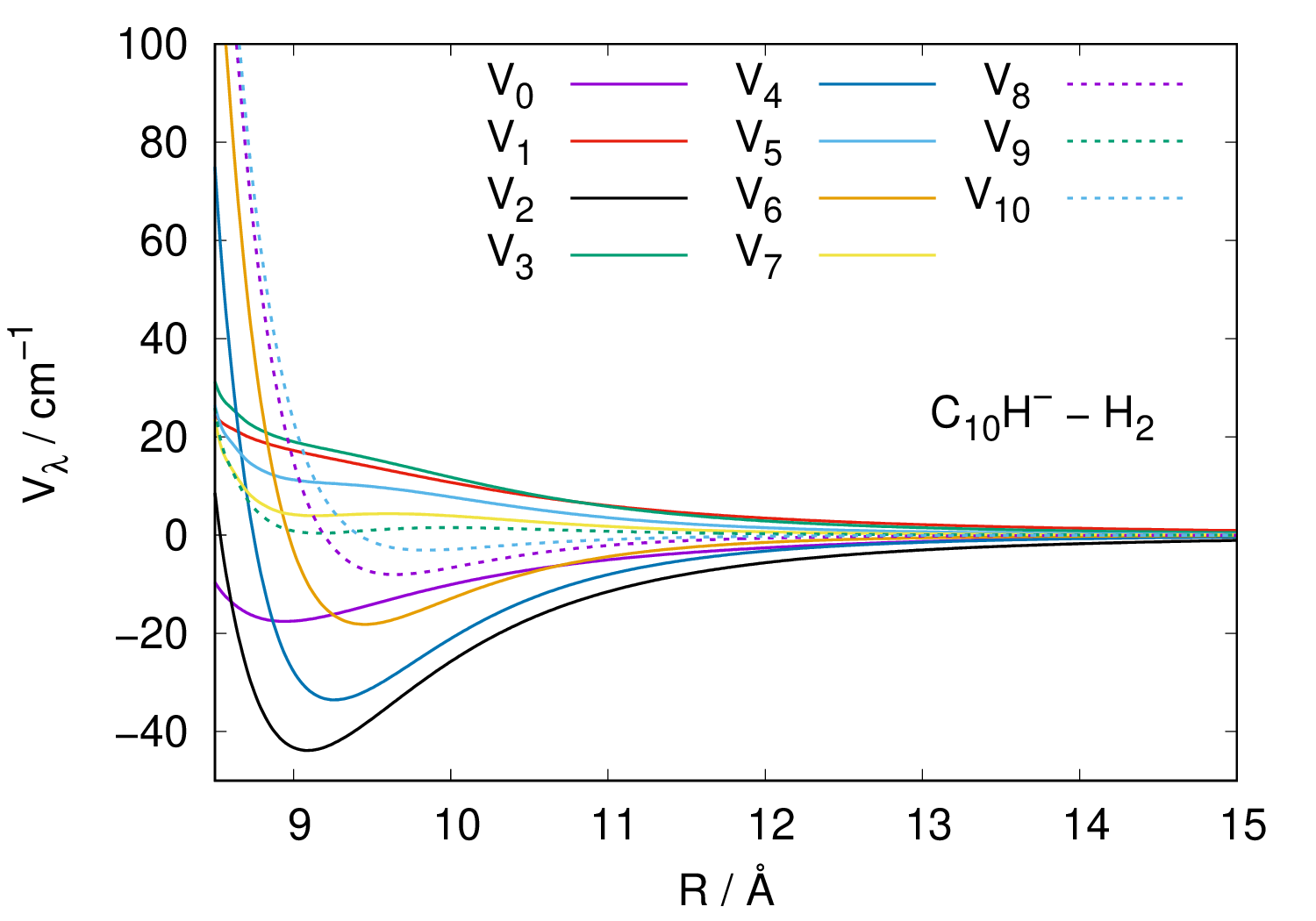} 
\caption{Plot of the Legendre coefficients as a function of $R$ for the   C$_{10}$H$^-$ - H$_2$ system in the 2D  dimensionality.The H$_2$ molecule is treated as a structureless partner. The 2D-PES is derived from the  average of  combinations of $\alpha$ and $\beta$ angles as discussed in the main text.} 
\label{fig7}
\end{figure}

   The procedure which we have followed in this case has been the accurate numerical expansion reported by equation (\ref{eq:vl}), whereby the same large number of multipolar radial coefficients indicated for the  C$_7$N$^-$$-$H$_2$ system have been generated. The lower, more important terms of that expansion are reported in Figure \ref{fig7}.

It is interesting to note here, given the differences in the shape of the odd-labelled terms in Figures 5 and 7, that we have used in fact  different scales. The V$_{\lambda}$ values go up to 300 cm$^{-1}$ in Figure 5 and only up to 100 cm$^{-1}$ in Figure 7. The R axis starts from 5.5 \AA \ in Figure 5, but from 8 \AA \ in Figure 7. Further, C$_{10}$H$^-$ is a longer molecule than C$_7$N$^-$, so that  the former is more prolate than the latter. The former longer chain is also more symmetric, despite the H atom present at one end. It therefore follows that for C$_{10}$H$^-$ the odd-labelled multipolar  terms are smaller in magnitude than the even-labelled terms, this occurring much more prominently than for the  data involving C$_7$N$^-$.

One sees again that the dynamical coupling induced by this interaction shows once more the dominance of the coefficients with ${\lambda}$= 2 and = 4, as we had found for the previous cyanopolyyne anion, while however the onset of the repulsive walls are located farther out at larger distances, as predicted by the longer linear structure of this anionic target.
The long range potential is the same as given in equation (\ref{eq:lr}) except that $\mu$ = 14.265 D for C$_{10}$H$^-$. Specifically, the switching function employed for the C$_{10}$H$^-$- H$_2$ system has been the same as the one reported by equations (\ref{eq:vf}) and (\ref{eq:sf}) using an $R_0$ value of 15 \AA\ and a value of $\Delta R$ = 0.5 \AA\ as switching parameters. The LR part is the same as the one reported by equation (\ref{eq:lr}) above, using the specific dipole moment value for  the longest polyyne anion of interest here.

\subsection{Treatment  of the Quantum Dynamics }

The standard time-independent formulation of the Coupled-Channel (CC) approach to quantum  scattering has been known for many years already
(see for example \citet{Taylor2006} for a general text-book formulation) while the more recent literature on the actual computational methods has been also very large. For an indicative set of references over the  years see for instance: \citep{60ArDaxx, 79Secrxx, 97KoHoffxx, 94JmHxx, 79FaGxx}. However, since we have already discussed our specific computational methodology in many of our earlier 
publications \citep{03MaBoGi, 08LoBoGi, 15GoGiCa}, only a short outline of our approach will be given in the present discussion.

For the case where no chemical modifications are induced in the molecule by the impinging projectile, the total scattering wave function can be
expanded in terms of asymptotic target rotational eigenfunctions (within the rigid rotor approximation) which are taken to be spherical
harmonics and whose eigenvalues are given by $B_{\textnormal{e}}j(j+1)$, where $B_{\textnormal{e}}$ is the rotational constant  mentioned already in the previous Section. The channel components for the CC equations are therefore expanded into products of total angular momentum
$J$ eigenfunctions and of radial functions to be determined via the solutions of the CC equations \citep{03MaBoGi, 08LoBoGi} i.e. the 
familiar set of coupled, second order homogeneous differential equations:
\begin{equation}
\left(\frac{d^2}{dR^2} + \mathbf{K}^2 - \mathbf{V} - \frac{\mathbf{l}^2}{R^2} \right) \mathbf{\psi}^J = 0.
\label{eq:CC}
\end{equation}

In the above coupled equations, the $\mathbf{K}^2$ matrix contains the wavevector values for all the coupled channels of the problem and the $ \mathbf{V}$ matrix contains the full matrix of the anisotropic coupling potential.  The required scattering observables are obtained in the asymptotic region where the Log-Derivative matrix has a known form in terms of free-particle
solutions and unknown mixing coefficients. Therefore, at the end of the propagation one can use the Log-Derivative matrix to obtain the 
K-matrix by solving the following linear system:
\begin{equation}
(\mathbf{N}' - \mathbf{Y}\mathbf{N}) = \mathbf{J}' - \mathbf{Y}\mathbf{J}
\end{equation}
where the prime signs indicate radial derivatives, $\mathbf{J}(R)$ and $\mathbf{N}(R)$ are matrices of Riccati-Bessel and Riccati-Neumann functions \citep{08LoBoGi}. The matrix  $\mathbf{Y}(R)$ collects the eigensolutions along the radial region of interest, out of which the Log derivative matrix is then constructed \citep{08LoBoGi}.
From the K-matrix produced by solving the coupled radial equations the S-matrix is then easily obtained and from it the state-to-state cross sections \citep{08LoBoGi}. We have already published an algorithm that
modifies the variable phase approach to solve that problem, specifically addressing the latter point and we defer the interested reader to
that reference for further details \citep{03MaBoGi,08LoBoGi}.

Because of the specific structural properties of the two title anions, both systems exhibit fairly small rotational constants, with the $B_{\textnormal{e}}$ value for C$_7$N$^-$ being 0.019436 cm$^{-1}$ and that for C$_{10}$H$^-$ being 0.010054 cm$^{-1}$. Hence, even at the lowest collision energies of interest in this study it is  necessary to couple within the quantum dynamics a rather large number of rotational states  to converge the Coupled Channel expansion, as further discussed below.
It therefore follows  that to carry out exact  CC calculations \citep{Green} for such systems would be computationally highly time-consuming. Therefore, we decided to carry out  some exact CC calculations in the low-energy range but switch to a less costly scheme i.e. to the Helicity Decoupled (HD) quantum scattering approximation, at the remaining energies for both systems \citep{kouri1975decoupling}. To be able to compare the results for the two systems under study, we have used the HD approximation also for the simpler case of the  C$_5$N$^{-}$/ H$_2$ system in our earlier work \citep{LGS2023} and shown that the results of the HD calculations for these linear systems turn out to be very close to those from the exact CC calculations, especially as the collision energy increases.

 While the CC calculation uses the space-fixed frame, the HD approximation uses the rotating body-fixed frame in which the projection quantum number ($K$) for the rotational angular momentum (\textbf{j}) of the individual rotor (or the collective \textbf{j}$_{12}$ in the case of two rotors) is taken to be with respect to the interparticle axis $R$. While the interaction matrix is diagonal in $K$, the \textbf{L}$^2$ ( = (\textbf{J-j})$^2$ operator, where \textbf{J} is the total angular momentum) is non-diagonal in $K$. The HD approximation neglects the off-diagonal matrix elements in $K$, but evaluates the diagonal matrix elements exactly, as given in the following equation 

\begin{equation}
      <jKJ|L^2|jKJ> = J(J+1) + j(j+1) - 2K^2
\end{equation}

It is important to point out that in this approximation only basis functions with $K \le j$ are included in the expansion set. For a detailed discussion on decoupling angular momenta in molecular collisions, the reader is referred to the work of Kouri \citep{kouri1975decoupling}. For an example of a successful use of the HD approximation for a tri-atom system, the reader is also referred to our earlier work of Buonomo et al. \citep{Buonomo}.

We have computed the rotational inelastic cross section ($\sigma$) values for a range of $E_{\textrm{coll}}$ and the rate coefficient ($k$) values for a range of temperatures between $T$ = 10 and 50 K. It turns out that the range of $E_{\textrm{coll}}$ and $T$ investigated allows us to make a significant comparison of our computed results with the results reported earlier \citep{Lara-Moreno2017, Lara-Moreno2019} for a similar, but smaller, molecular ion: the C$_3$N$^{-}$. 

We further found that, in practice, the presently  obtained results can be reliably extrapolated to higher temperatures, although for a safer comparison we also carried out the scattering calculations for both systems up to 400 cm$^{-1}$, thus allowing us to obtain converged rate coefficients up to 50 K (see the discussion below).

To specifically test the numerical effects of using either the HD or the CC treatments, we compare in Figure 8 the two methods for a selected set of the many transitions discussed below and involving here the longer anion- We are also reminding the reader that below the 150 cm$^{-1}$ energy value we used only the CC results, leaving the use of HD decoupling to the calculations above that limit.

\begin{figure}
\centering
\includegraphics[width=1.0\linewidth,angle=+0.0]{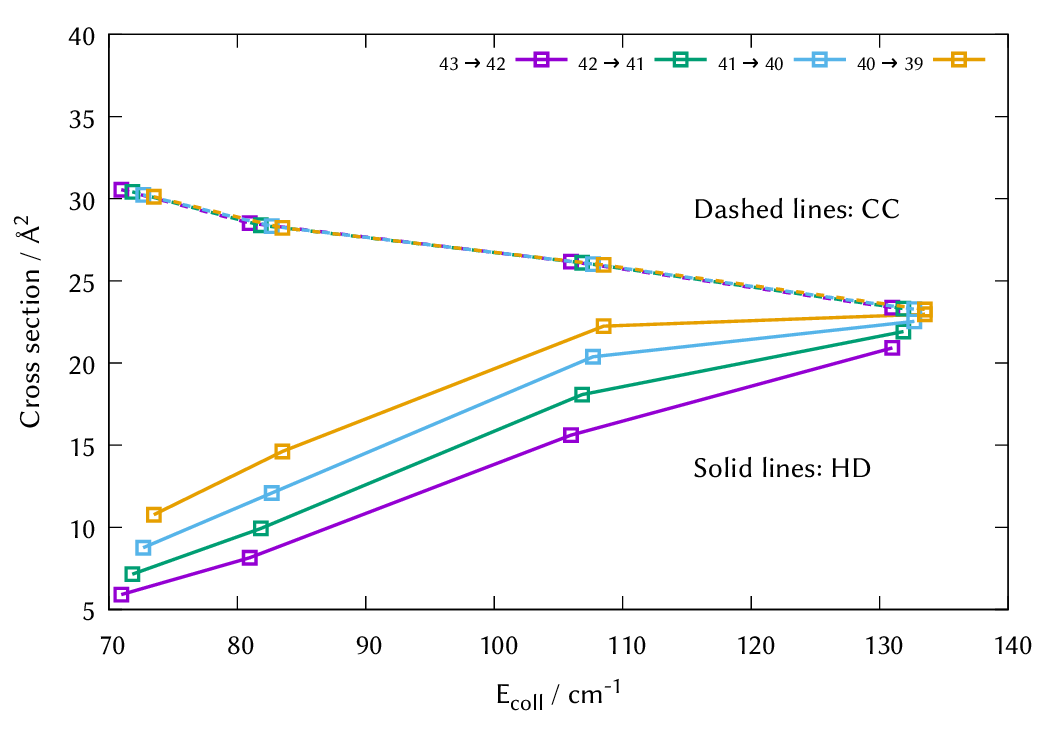}
\caption{Comparison between Cross section values computed using the CC coupling scheme (dashed lines) or the HD decoupling approximation (solid lines) for the longer anion C$_{10}$H$^-$ in collision with H$_2$. See main text for further details.}
\label{fig8}
\end{figure}

We clearly see from the results shown in that Figure 8 that, while the HD calculations differ from the CC calculations at the lower energies, at collision energies around 150 cm$^{-1}$ the two methods provide essentially the same values of cross sections, the very small differences being due to the differences in the coupling of rotational angular momenta in the case of the HD calculations (see before). It therefore follows that it makes sense to employ either of the coupling schemes by using the energy ranges below and above the 150 cm$^{-1}$ as a separation for choosing one or the other, as illustrated by the data in Figure 8.

The inelastic rate coefficients $k_{j \to j'}(T)$ were obtained using the standard numerical integration procedure as the convolution of the cross sections over a Boltzmann distribution of $E_{\textrm{coll}}$  (all in atomic units): 

\begin{equation}
k_{j \to j'}(T) \,=\, \left(\displaystyle \frac{8}{\pi \mu k_{B}^3 T^3 }
\right)^{1/2}
  \int_0^{\infty} \sigma_{j \to j'}(E_{\textrm{coll}})
E_{\textrm{coll}} \, e^{-E_{\textrm{coll}}/k_{B}T} \, 
\rm{d} E_{\textrm{coll}}
\label{eq.rateK}
\end{equation}

The numerical integration was checked for the convergence of the final rate coefficient values at different points of the required range of temperatures. It is also interesting to note that the values of the reduced mass appearing in the denominator in eq.(9) are given (in amu) by very close values for the two anions: 1.97503 for C$_7$N$^{-}$ and 1.9829 for C$_{10}$H$^-$. This suggests that the observed differences in size between the rates are largely controlled by the differences in size of their corresponding inelastic cross sections and guided by differences within their coupling PESs.

All the scattering calculations for the C$_7$N$^{-}$ anionic partner were carried out exactly using the CC treatment. For the longer chain of the C$_{10}$H$^-$ anion we also used the exact CC coupling scheme at the lower collision energies, while switching to the HD calculations at the higher energies from 150 cm$^{-1}$. All calculations were carried out using  the MOLSCAT \citep{MOLSCAT2,MOLSCAT} computer code.

Since the working of the MOLSCAT code and the theory behind it have been already extensively described in the literature \citep{MOLSCAT2,MOLSCAT}, we only mention here the parameters used in the current study. For the case of the H$_2$ partner, treated as a point mass, interacting with the shorter molecular anion C$_7$N$^-$, we employed the following data: JTOTL = 0, JTOTU = 80$-$100 depending upon $E_{\textrm{coll}}$, BE = 0.019436, JMAX = 60, LMAX = 40. RMAX was taken to be as large as 500 \AA\ down to the lowest energy ($E_{\textrm{coll}}$ = 0.0001 cm$^{-1}$) collisions and it was lowered to a value of 200 \AA\ for the higher energies we have considered. We remind the reader once more that our inelastic cross sections were generated up to collision energy values of 400 cm$^{-1}$.

For the longer linear chain (C$_{10}$H$^-$) interacting with a point mass H$_2$  discussed in the present work,  the parameters used are: JTOTL = 0, JTOTU = 80, varying  up to 119 for $E_{\textrm{coll}}$= 125 cm$^{-1}$, BE = 0.010054,  JMAX = 70$-$90 depending upon $E_{\textrm{coll}}$,  LMAX = 95, while RMAX was taken to be as large as 300 \AA\ for lower range of  energy ($E_{\textrm{coll}}$ = 0.0001$-$100.0 cm$^{-1}$) collisions and it was lowered to a value of 50 \AA\ for $E_{\textrm{coll}}$ = 100$-$400 cm$^{-1}$.

\section{Results and Discussion}

\subsubsection{Comparing  C$_7$N$^-$$-$H$_2$ rate coefficients  with those for shorter cyanopolyynes.\label{sec:3.1}}

\begin{figure}
\centering
\includegraphics[width=1.0\linewidth,angle=+0.0]{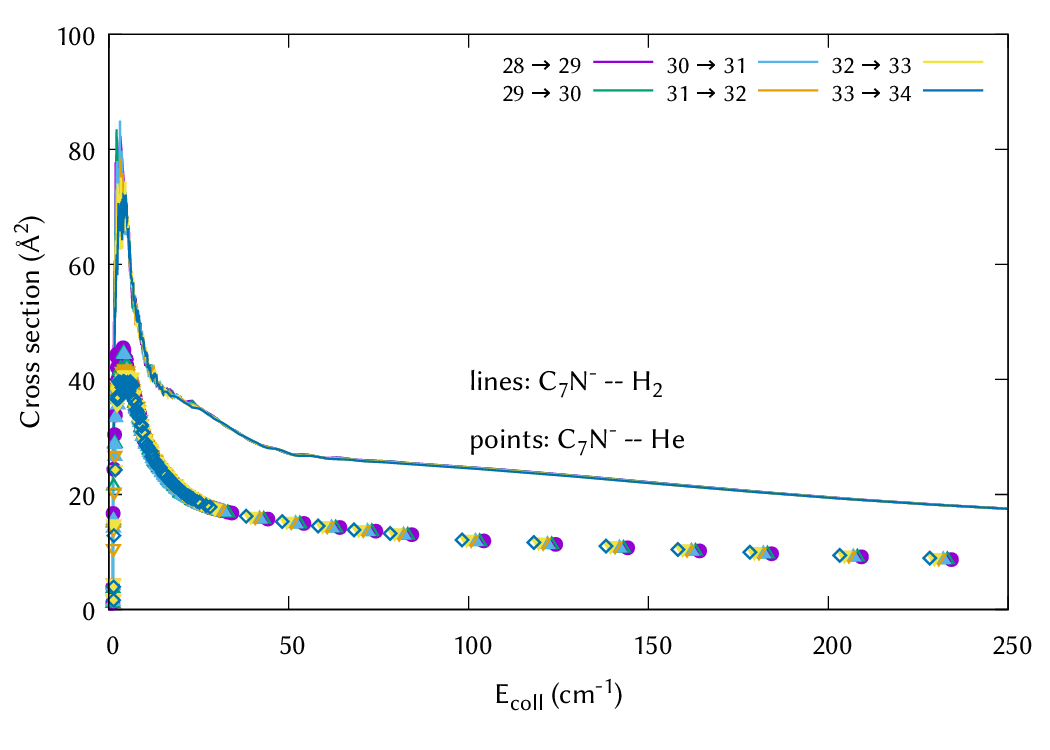}
\caption{Excitation cross section values plotted as a function of $E_{\textrm{coll}}$ for a variety of initial rotational levels  of  C$_7$N$^-$ in collision with H$_2$ treated as a point mass. The results of our earlier calculations \citep{LGS2023} with He as a collision partner are also included for comparison (diamond curves). See main text for further details.}
\label{fig9}
\end{figure}

 To verify once again the existing size differences between the state-changing cross sections generated with the He partner and those produced with H$_2$, we show in Figure \ref{fig9} a comparison between the present results (solid lines) and the earlier ones for the He projectile \citep{LGS2023} (diamond curves).

To start with, it may be noted that the cross section values for $\Delta j$ = 1 transitions from several initial $j$ states are comparable in magnitude and their $E_{\textrm{coll}}$-dependence is nearly indistinguishable because of the small $B_\textnormal{e}$ value for the anionic partner. This is not surprising because the energy gap for the different transitions ranges from 1.126 cm$^{-1}$ (28-29) to 1.28 cm$^{-1}$ (33-34) only.   

\begin{figure}
\centering
\includegraphics[width=1.0\linewidth,angle=0.0]{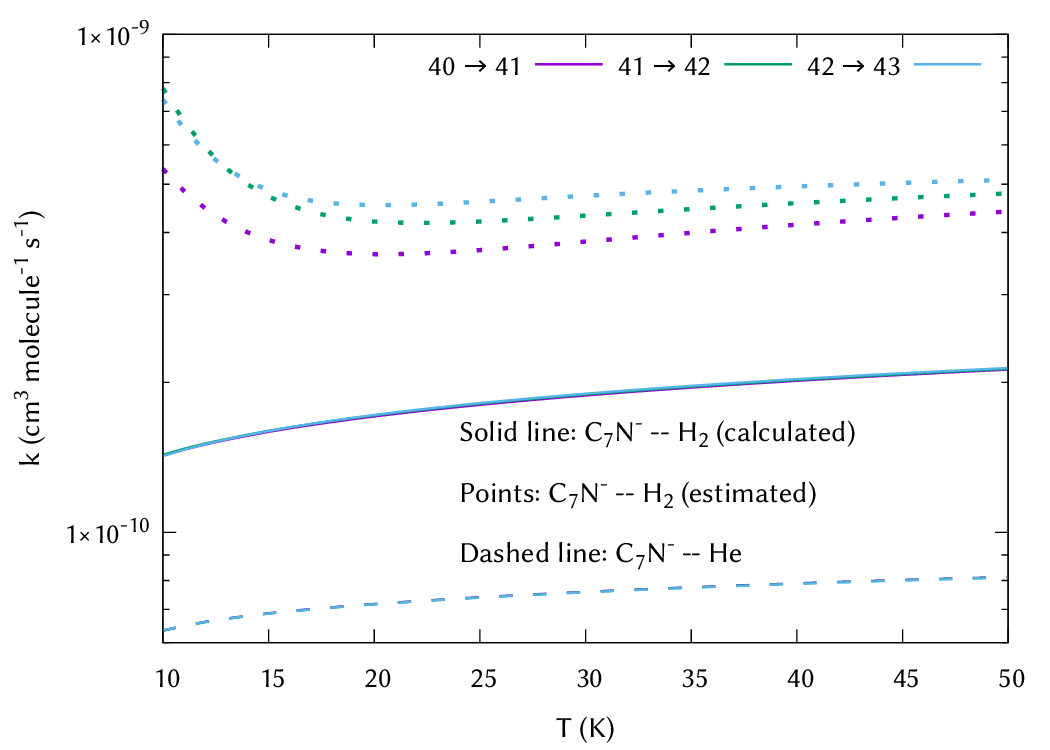}
\caption{Excitation rate coefficients for a sampling of rotational transitions involving C$_7$N$^-$ in collision with H$_2$ treated as a point mass using our present orientation-averaged, 2D potential. The solid lines represent the results of present calculations while the dotted lines report the estimated values based on what was given in \citep{LGS2023}. The $k$ values for the same set of rotational transitions but for He as a collision partner are given by the dashed lines at the bottom of the figure. See main text for further details.}
\label{fig10}
\end{figure}

It can be seen further  from Figure \ref{fig9} that the cross section values for the excitation processes when H$_2$ is the partner of the cyanopolyyne are markedly larger than for the case of the He projectile. They are the largest near threshold energies (where the increase factor is  about 2.5) and remain a factor of about 1.5 larger all the way up to the largest collision energy investigated. This is in line with what we had suggested in our earlier work \citep{LGS2023}, and is further confirmed by the comparison between excitation rate coefficients shown in  Figure \ref{fig10} for slightly different rotational transitions. We see there that calculated rate coefficients using the 2D approximation confirm that they are larger than those produced by He as a collision partner. Furthermore, we see that the rates estimated by a scaling procedure described in our earlier work\citep{LGS2023} (dotted lines) are on the whole a factor of two too large with respect to the accurate calculations. That procedure essentially took advantage of the fact that we knew the correct scaling between the two types of rates for the case of the smaller anion C$_5$N$^{-1}$ and therefore we used it to scale the rates for the present C$_7$N$^{-1}$ anion. Such a procedure is shown by the present calculations not to be accurate enough for being used reliably.

The results for inelastic rate coefficients shown in  Figure \ref{fig10} report calculations done using the present 2D PES for a selection of transitions which had been cited  in the observational sighting of this anion \citep{Cern2023}. One clearly sees that our computed rate coefficients are indeed fairly large and are of the order of about 2x10$^{-10}$ cm$^3$ molecule$^{-1}$s$^{-1}$. Our earlier estimates from a scaling of the same processes from the shorter  C$_5$N$^{-}$ were discussed in \citep{LGS2023} and are also reported here by the three dotted curves. They turn out to be too large when compared to the correct calculations  from the present study. As is to be expected, the corresponding rates for He as a collision partner are about one order of magnitude smaller and are given by the dashed curves at the bottom of the Figure \ref{fig10}. The same general behaviour is also shown by the set of de-excitation processes reported in Figure 11. It should also be noted here that we are consistently looking at transitions between highly excited rotational levels which carry very large degeneracy factors that change little between excitation and de-excitation processes. Furthermore, given the very small energy gaps between such transitions, it stands to reason that the results for the two types of processes appear largely to be of the same size in  the figures we are reporting here.

\begin{figure}
\centering
\includegraphics[width=1.0\linewidth,angle=0.0]{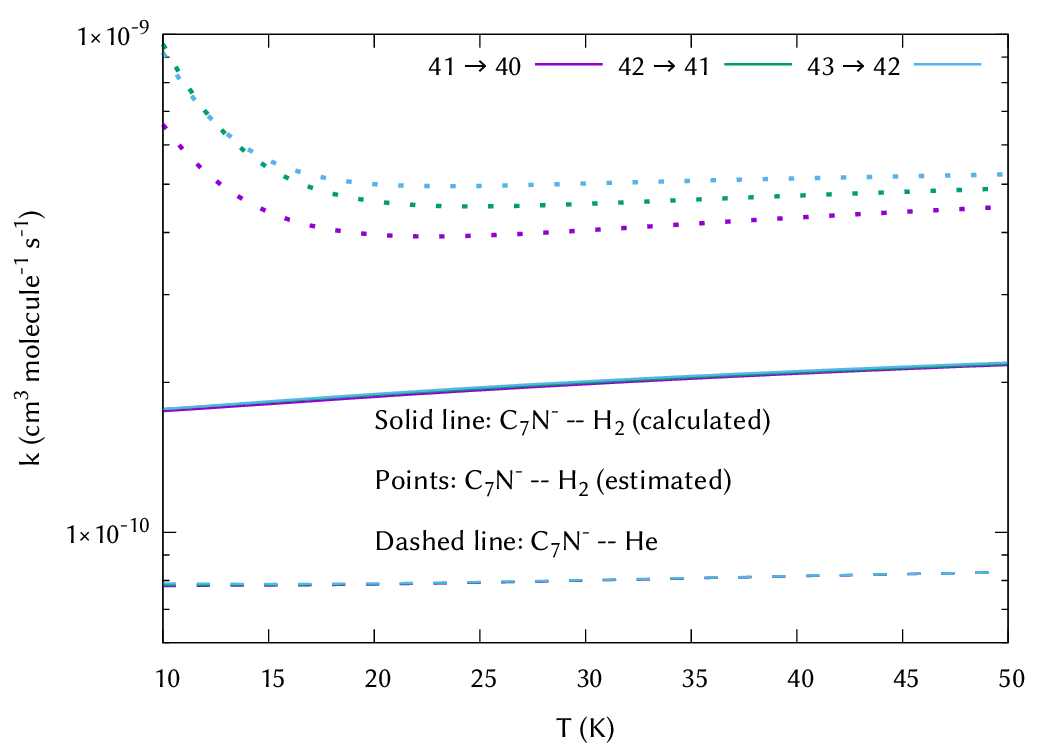}
\caption{De-excitation rate coefficients for the pairs of rotational states of C$_7$N$^-$ in collision with H$_2$/He presented in Figure \ref{fig10}. See main text for further details.}
\label{fig11}
\end{figure}

 What we see here confirms that the longest cyanopolyyne anion observed thus far has very large rate coefficients involving its rotational state-changing processes from collision with the H$_2$ partner. They already indicate that the  longer the  chain system, the larger are expected to be the associated  state-changing collision rate coefficients with H$_2$ as the collision partner.

  \begin{figure}
\centering
\includegraphics[width=1.0\linewidth,angle=0.0]{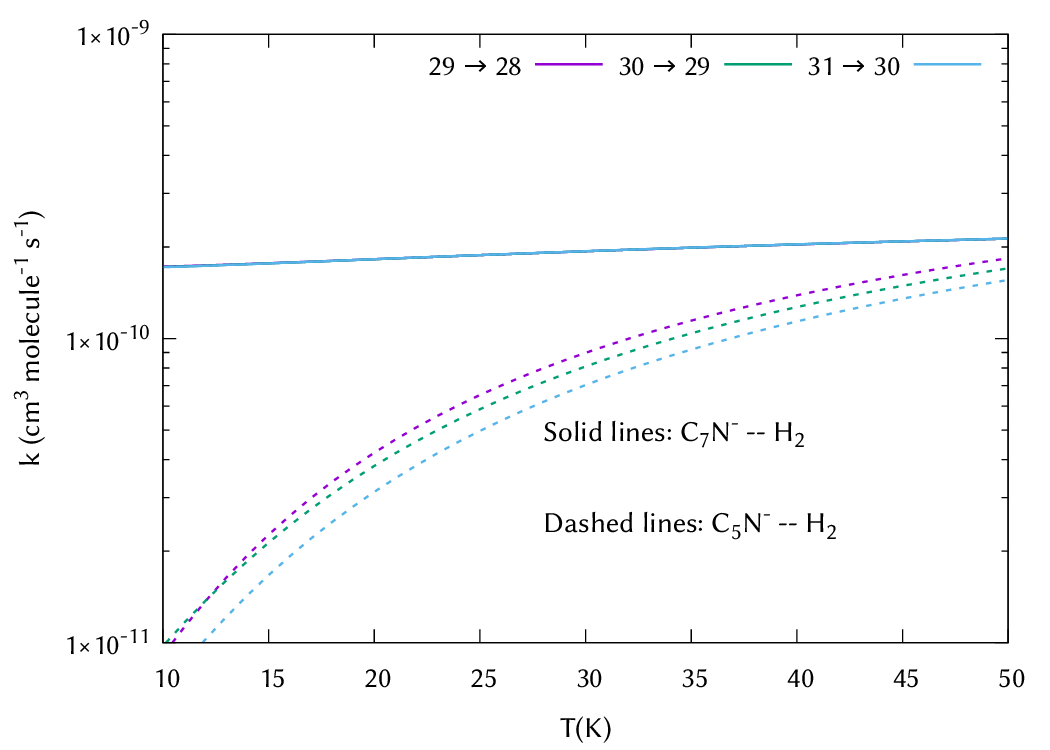}
\caption{Comparing de-excitation rate coefficients for three of the observed transitions involving C$_7$N$^-$ (solid lines) and C$_5$N$^-$ (dashed lines) in   collision with H$_2$. See main text for further details.}
\label{fig12}
\end{figure}

\begin{figure}
\centering
\includegraphics[width=1.0\linewidth,angle=0.0]{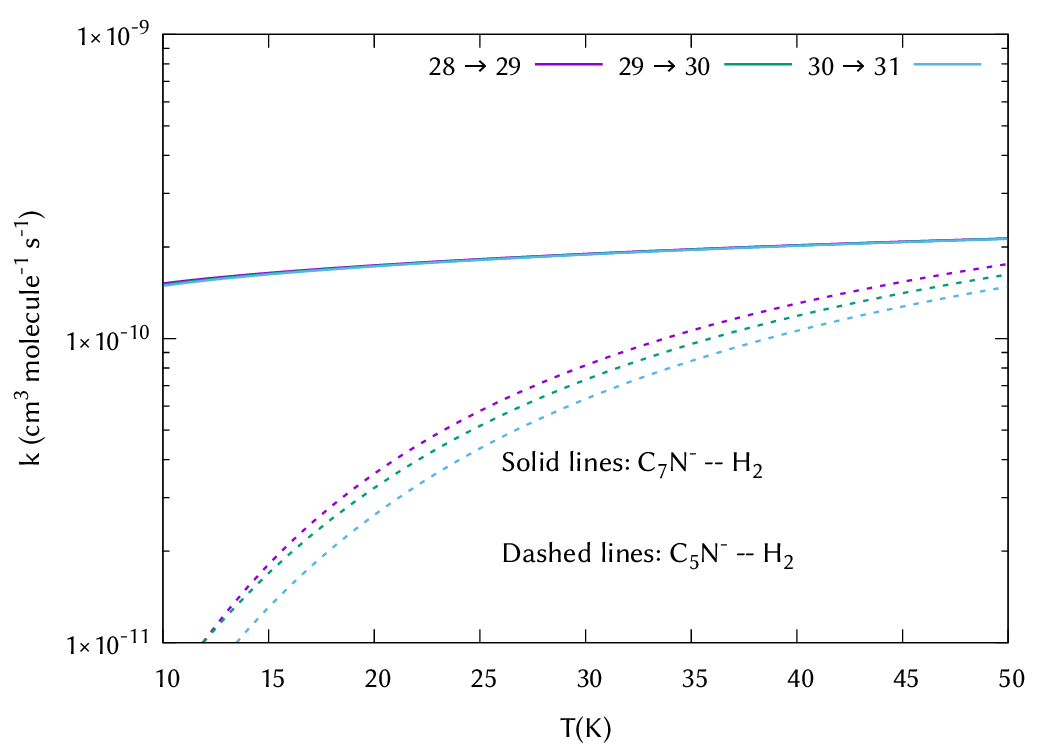}
\caption{Comparison of excitation rate coefficients for complementary rotational transitions in two different cyanopolyynes in collision with H$_2$ reported in Figure \ref{fig11}. See the main text for further details.}
\label{fig13}
\end{figure}

The state-changing  processes reported in Figures \ref{fig12} and \ref{fig13} carry out a different comparison: one involving  the de-excitation transitions (Figure \ref{fig12}) and excitation processes (Figure \ref{fig13}) of the longest detected cyanopolyyne and the next shorter term in the series of chains: the  C$_5$N$^-$, also in collision with H$_2$ and discussed in an earlier work of our group \citep{BGG23}. The excitation rate coefficients shown in Figure \ref{fig13} depict the results for the C$_5$N$^-$ via a series of dashed curves while the present results for the longer anion are given by the solid lines. The rate coefficients for C$_5$N$^-$ are about one order of magnitude smaller than those pertaining to the longer chain. If we note here that the rotational constants, and hence the corresponding energy spacing between rotational states, vary from 0.046292 cm$^{-1}$ for the C$_5$N$^-$ down to 0.019436 cm$^{-1}$ for the C$_7$N$^-$, we can see that a higher density of states for the latter anion yield  more efficient energy transfer probabilities  by collision than the former, shorter term of the series. This type of behaviour is further confirmed by the data in Figure \ref{fig12}, where the de-excitation rate coefficients are compared: the state-changing collision probabilities for the  C$_5$N$^-$ anion are once more markedly smaller than those pertaining to the longer cyanopolyyne.

 \begin{figure}
\centering
\includegraphics[width=1.0\linewidth,angle=0.0]{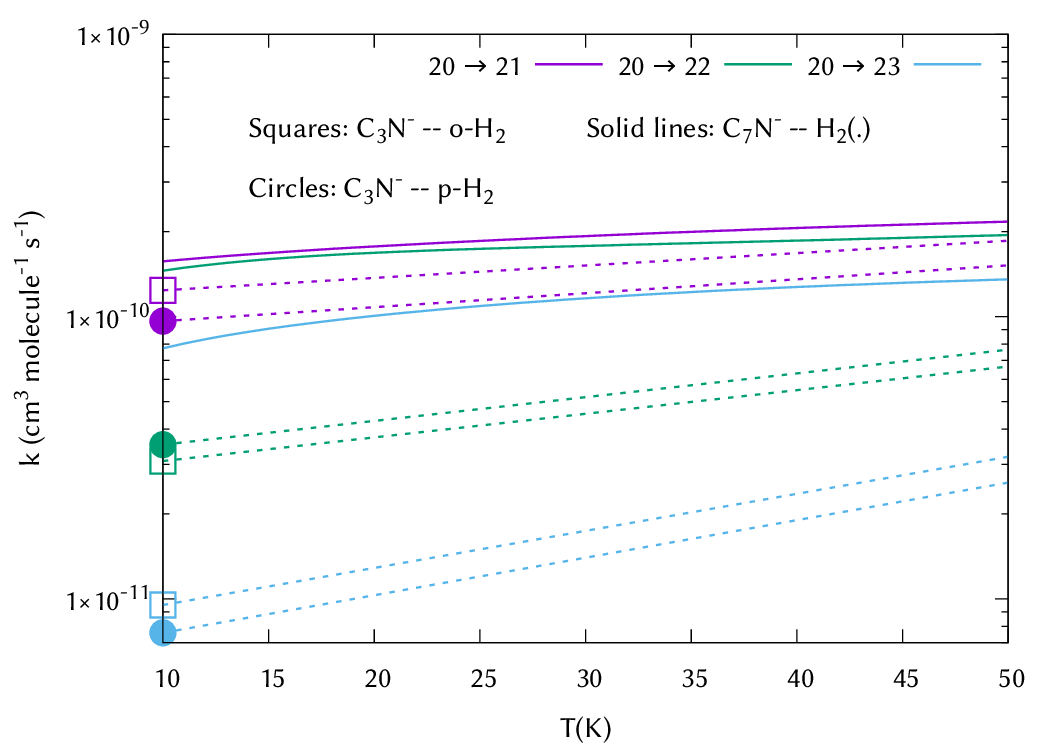}
\caption{Comparing excitation rate coefficients for three of the observed transitions involving C$_7$N$^-$ (solid lines) and C$_3$N$^-$ (dashed lines) in   collision with H$_2$. The data for the C$_3$N$^-$ are taken from the earlier work of Moreno et al.  \citep{Lara-Moreno2019}. See main text for further details.}
\label{fig14}
\end{figure}

 \begin{figure}
\centering
\includegraphics[width=1.0\linewidth,angle=0.0]{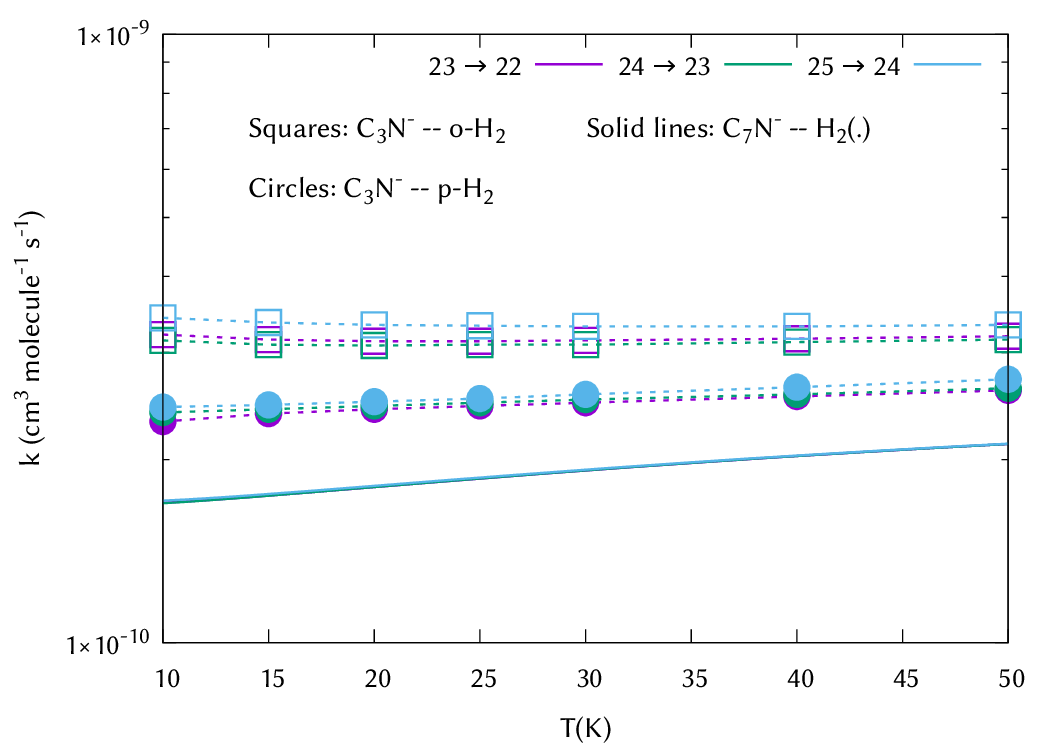}
\caption{Comparing de-excitation rate coefficients for three of the observed transitions involving C$_7$N$^-$ (solid lines) and C$_3$N$^-$ (dashed lines) in   collision with H$_2$. The results for the smaller anion are from the earlier work in ref. \citep{ Lara-Moreno2019}. See main text for further details.}
\label{fig15}
\end{figure}

   An additional set of comparison is reported by the data in Figures \ref{fig14} and \ref{fig15}, where we now investigate the differences between the state-changing collision processes of one of the present cyanopolyyne  anions (C$_7$N$^-$) and those for yet another term of the same series, the C$_3$N$^-$ anion, for which calculations had been reported in earlier work by the Bordeaux group  \citep{ Lara-Moreno2019}.
   The transitions are now  between  different sets of levels, with the rotational constant of this shorter molecule being 0.1618 cm$^{-1}$, with its permanent dipole moment of 3.10 Debye. The marked changes in the energy gaps between levels when going from the shorter anion (C$_3$N$^-$) to the longer one are once  strongly affecting the efficiency of collision energy transfer between rotational states. We see,in fact, from the data in Figure \ref{fig14} that the rate coefficients for the excitation processes involving the C$_3$N$^-$ partner of H$_2$ are smaller by up to about  a factor of five when compared to those we obtained for  the  C$_7$N$^-$. On the other hand, the de-excitation processes in Figure \ref{fig15}, show that the inelastic rate coefficients for the C$_3$N$^-$ become even slightly larger than those for the  C$_7$N$^-$. This result is at variance with those  in Figure \ref{fig14} and with all the similar results for the other systems discussed by Figures from 9 to 12. This odd behaviour, however, might be due to the fact that Lara-Moreno et al. \citep{ Lara-Moreno2019} used the $J$-shifting approximation in their calculations while we are employing the exact CC coupling scheme, as discussed earlier. It is therefore hard to decide how much to trust the approximate calculations with respect to the accurate ones.

\subsubsection{  Computed cross sections and rate coefficients for the C$_{10}$H$^-$$-$H$_2$ and a comparison with C$_7$N$^-$$-$H$_2$} 

The C$_{10}$H$^-$  anion is the longest chain of the linear (C,H)-bearing anions which have been detected so far in similar ISM environments. It is also the system with the smallest rotational constant (0.010054 cm$^{-1}$), a feature implying a very large number of rotational channels which have to be coupled during the scattering calculations. Hence  our present results reported below originate from very time-consuming scattering calculations for the C$_{10}$H$^-$$-$H$_2$ system.

\begin{figure}
\centering
\includegraphics[width=1.0\linewidth,angle=+0.0]{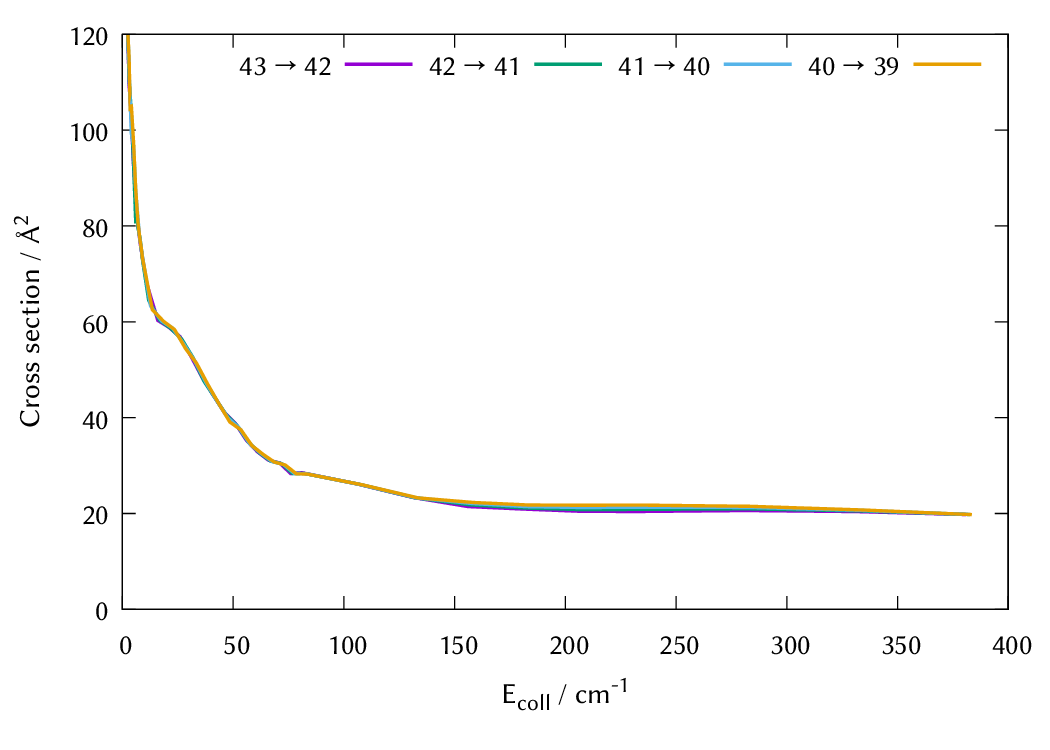}
\caption{Relaxation cross section values plotted as a function of $E_{\textrm{coll}}$ for a variety of initial rotational levels  of  C$_{10}$H$^-$ in collision with H$_2$ treated as a point mass. See main text for further details.}
\label{fig16}
\end{figure}

The computed values of the de-excitation cross sections involving a broad range of transitions with the dominant $\Delta j$ = 1 selection rule are reported by Figure \ref{fig16}. It is interesting to observe that such inelastic processes involve fairly large cross sections at threshold, even larger than those shown by the C$_7$N$^-$$-$H$_2$ system with the data of Figure \ref{fig9}. On the other hand, given that the energy gaps of the transitions in Fig.16 are larger than those shown in Fig.9, owing to the smallness of the rotational constants as discussed earlier, the cross sections for all transitions turn out to be smaller (around 20 \AA$^2$ ) than was the case for those involving the shorter cyanopolyyne anion reported in Figure \ref{fig9}, where their average size is around 30 \AA$^2$ .

\begin{figure}
\centering
\includegraphics[width=1.0\linewidth,angle=+0.0]{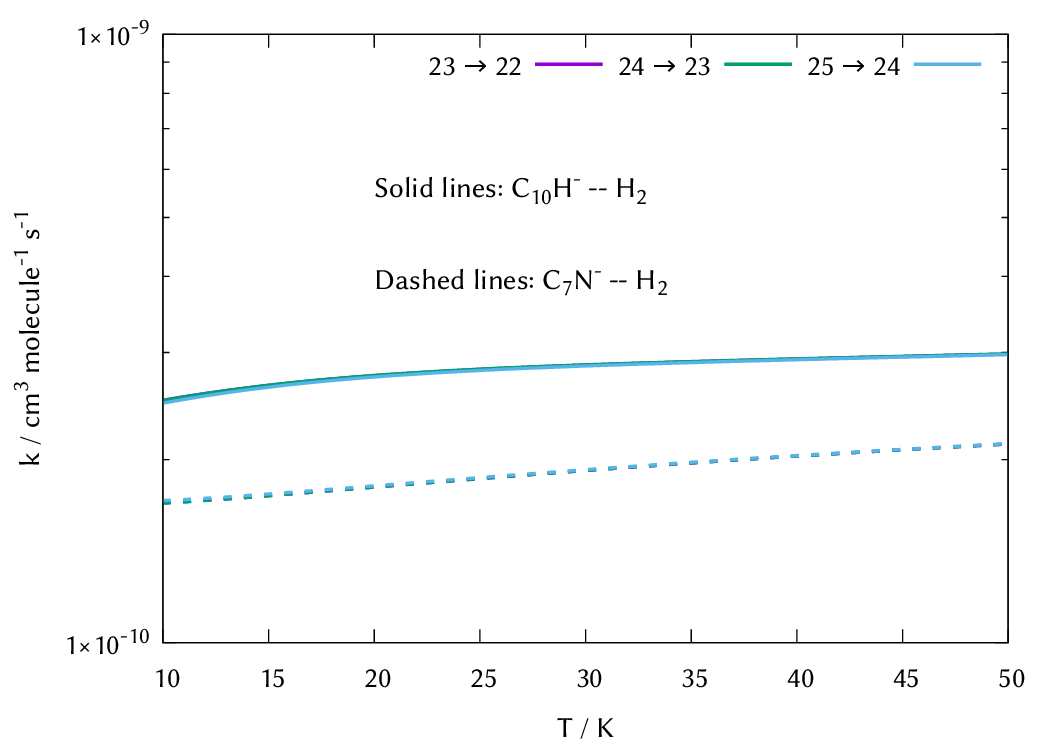}\caption{Comparison of relaxation rate coefficients  as a function of $T$ for a few of the rotational levels  associated with observations of  C$_7$N$^-$(see main text). The same levels are also presented for  C$_{10}$H$^-$, both systems in collision with H$_2$ treated as a point mass. See main text for further details.}
\label{fig17}
\end{figure}

\begin{figure}
\centering
\label{fig17a}
	\includegraphics[width=0.85\linewidth,angle=+0.0]{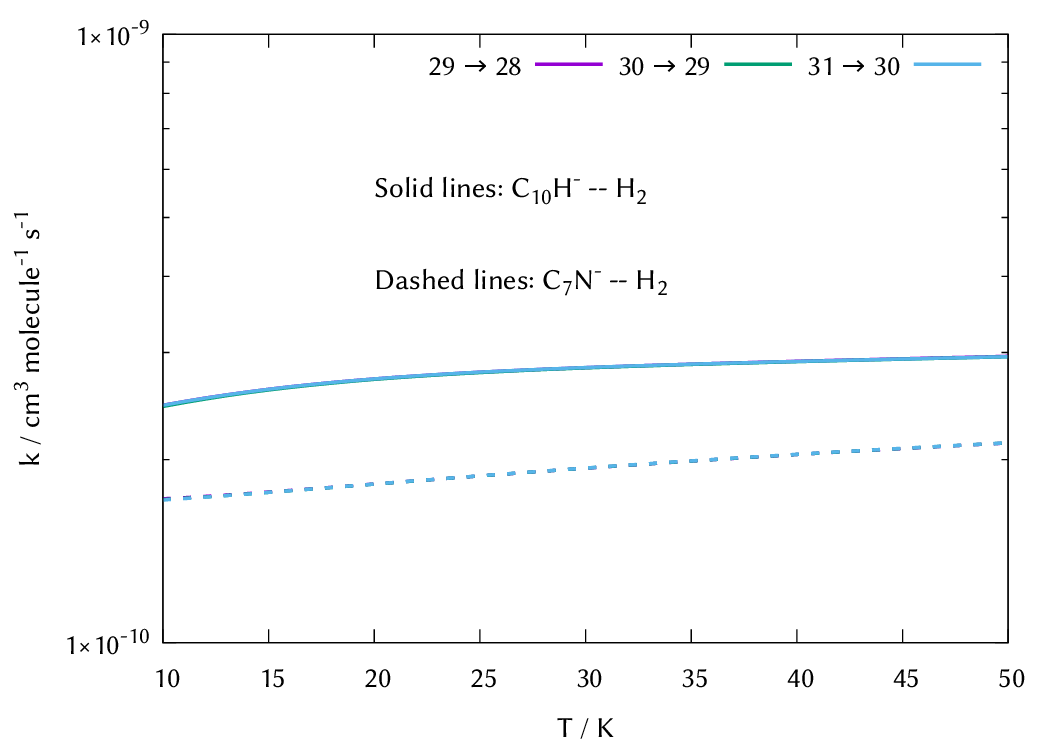} 
\label{fig17b}
	\includegraphics[width=0.85\linewidth,angle=+0.0]{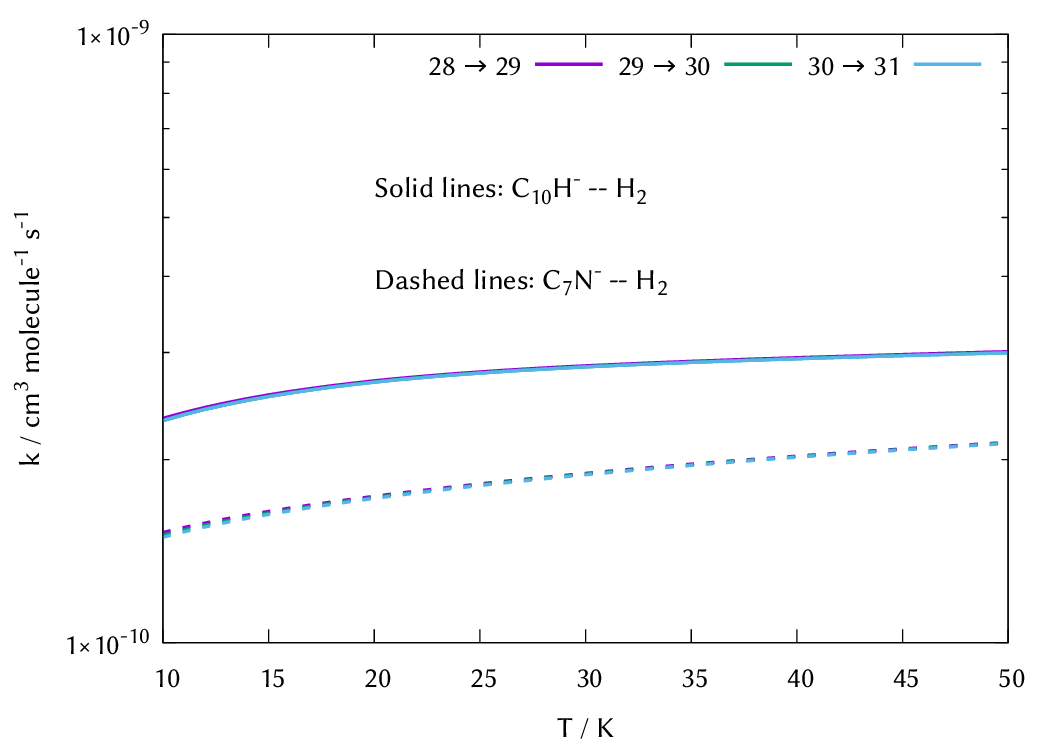}
\caption{Comparison of relaxation processes (upper panel) and excitation processes (lower panel) between computed rate coefficients for the C$_{10}$H$^-$ and C$_7$N$^-$ systems in collision with H$_2$ treated as a point mass. See main text for further details.}
\label{fig18}
\end{figure}

\begin{figure}
\centering
{\label{fig18a}
	\includegraphics[width=0.85\linewidth,angle=+0.0]{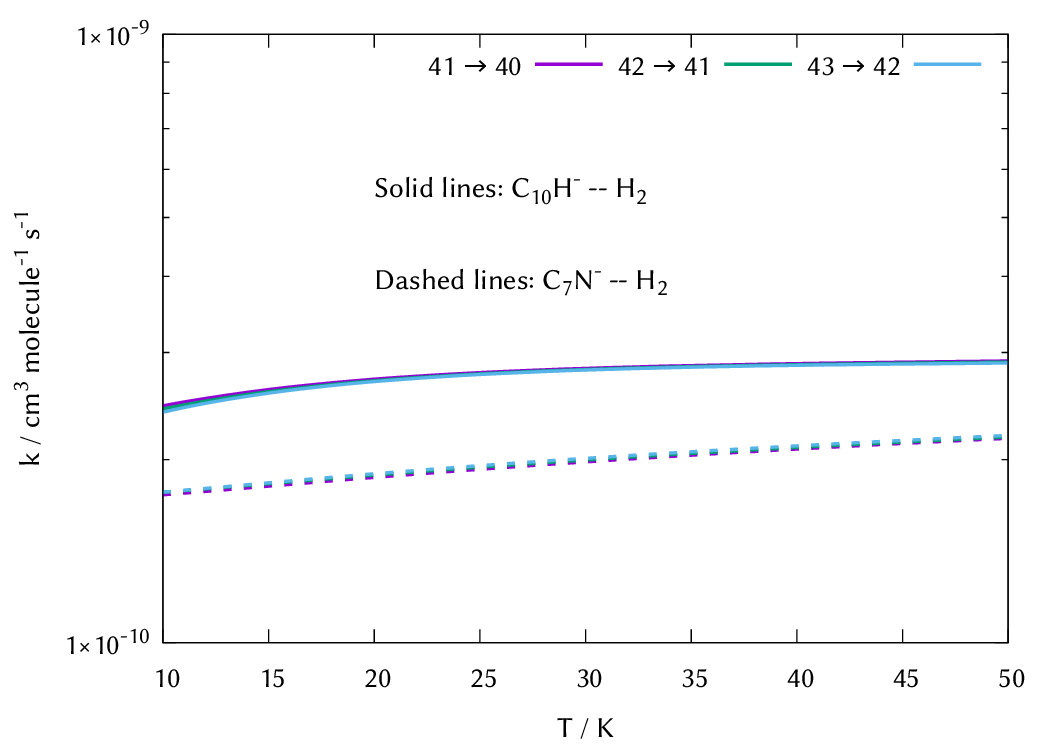}} 
{\label{fig18b}
	\includegraphics[width=0.85\linewidth,angle=+0.0]{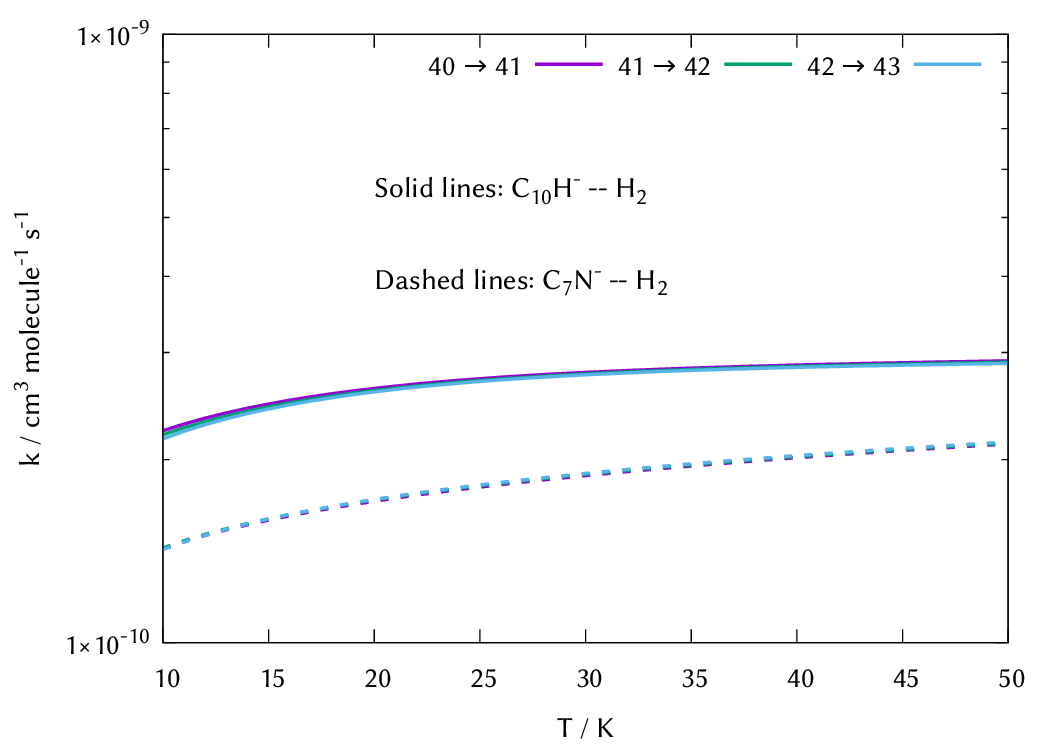}} 
\caption{Comparison of relaxation processes (upper panel) and excitation processes (lower panel) between computed rate coefficients for the C$_{10}$H$^-$ and C$_7$N$^-$ systems in collision with H$_2$ treated as a point mass. See main text for further details.}
\label{fig19}

\end{figure}

 The computed relaxation rates involving the rotational levels experimentally detected for C$_7$N$^-$, discussed in the Introduction Section and in ref.\citep{Cern2023}, are reported by the data in Figures \ref{fig17}, \ref{fig18} and \ref{fig19}, where they are compared with the same computed rate coefficients for the longer C$_{10}$H$^-$.

It is useful to make the following observations about the results shown in those figures:

(i) the de-excitation processes from the  levels for the shorter chain in Figure \ref{fig17} are reported on a semi-log scale to emphasize the fairly slow temperature dependence of these rates and their marked similarities in size and $T$-dependence. Furthermore, we notice that the ones from the longer chain (solid lines), which involve smaller energy gaps between transitions than those for the shorter anion (dashes), are  about a factor of two larger for all three transitions  and slowly increase with $T$. The same indeed occurs for the C$_7$N$^-$$-$H$_2$ system, where the relaxation rate coefficients are slightly smaller in size and increase very little with temperature;

(ii) when we look at the relaxation rate coefficients involving slightly higher rotational states, reported by the two panels in Figure \ref{fig18}, we see very much the same behaviour, both for the relaxation processes of the upper panel and the excitation transitions in the lower panel. The rate coefficients for the two linear chains differ by about a factor of two while we can verify once more the substantial similarities in  behaviour for this set of rates in both polyyne anions;

(iii) the computed rate coefficients finally given in the two panels of Figure \ref{fig19} involve  transitions between the highest rotational states considered in the present study. We see once more that their behaviour is essentially the same as in the two previous figures: very slow dependence on temperature for both systems, marked size similarities for the two anionic polyynes in collision with H$_2$ and the rates involving the C$_{10}$H$^-$ remaining larger by the same factor as in the previous figures. 

(iv) It is also interesting to note that excitation and de-excitation processes turn out to have comparable rate coefficients for the different transitions as anticipated in the earlier work that discussed a generalized  exponential gap law \citep{ PW72,GS80}:

\begin{equation}
P({j_i \rightarrow j_f}) = A\exp{(-C|\Delta E_{if}|/(k_\textnormal{B} T))}
\label{eq:gap_law}
\end{equation}

as both initial ($j_i$ ) and final ( $j_f$) states involved have nearly the same degeneracy and the energy gap values remain very similar in size along the considered energy scales.

On the whole, we also see that the computed state-changing rate coefficients for these two, fairly long anionic polyynes are consistently large and confirm that the systems exhibiting  smaller energy gaps in state-changing processes yield  larger rate coefficients when interacting with H$_2$.

 \subsubsection{Radiative Emission and Critical Densities.}  
 
  Another important process for characterizing the rotational internal state distributions of the anions  is their interaction
with the surrounding black-body radiation. A molecule interacts with such a field by absorbing
and emitting photons. The transition rates from an excited state $k$ can be written as sum of stimulated and spontaneous 
emission rates as \citep{03BrCaxx}:
\begin{equation}
\kappa^{em}_{k \to i} = \kappa^{sti}_{k \to i} + \kappa_{k \to i}^{spo} = A_{k \to i}(1 + \eta_{\gamma}(\nu,T))
\label{eq:tot_emi}
\end{equation}
where $A_{k \to i}$ is the Einstein coefficient for spontaneous emission and 
$\eta_{\gamma}(\nu ,T) = (e^{(h\nu/k_B T)} - 1)^{-1}$ is the Bose-Einstein photon occupation number.

The Einstein coefficient for dipole transitions is given as
\begin{equation}
A_{k \rightarrow i} = \frac{2}{3} \frac{\omega_{k \rightarrow i}^3 S_{k \rightarrow i}}{\epsilon_0 c^3 h (2 j_k + 1)}
\label{eq:dip_tran}
\end{equation}
where $\omega_{k \to i} \approx 2B_0(j_i + 1)$ is the transition's angular frequency, and $S_{k \rightarrow i}$ is the line strength,  a quantity clearly linked to the Einstein coefficients via the multiplicity factors pertaining to the present processes. The quantity $\epsilon_0$ corresponds to the vacuum permittivity in the Einstein equations. 

For pure rotational transitions, equation (\ref{eq:dip_tran}) simplifies to 

\begin{equation}
A_{k \rightarrow i} = \frac{2}{3} \frac{\omega_{k \rightarrow i}^3}{\epsilon_0 c^3 h} \mu_0^2 \frac{j_k}{(2 j_k + 1)}
\label{eq:dip_tran_rot}
\end{equation}
where $\mu_0$ is the internuclear electric dipole moment of the molecule.

The results of the present calculations are reported in Table \ref{tab:Ecoeffs}, where the parameters needed in equation (\ref{eq:dip_tran_rot}) are also explicitly given.

\begin{table}
    \centering
    \caption{Einstein Emission Coefficients (A$_{ij}$), in 10$^{-5}$ s$^{-1}$, for C$_3$N$^-$, C$_5$N$^-$, C$_7$N$^-$, and C$_{10}$H$^-$ with $\Delta j$=-1. The values of the B rotational constant in cm$^{-1}$, and the value of $\mu$ dipole moment in Debye, are provided in the table for each system.}
    \label{tab:Ecoeffs}
    \vspace{0.3cm}
    \begin{tabular}{l r r r r }
    \hline\hline\\[-0.15cm]
     & C$_3$N$^-$ & C$_5$N$^-$ & C$_7$N$^-$ & C$_{10}$H$^-$\\[0.2cm]
    $B_\textnormal{e}$ [cm$^{-1}$] &  0.161800 & 0.046292 & 0.01941 & 0.010054 \\
    $\mu$ [Debye] & 3.100 & 5.200 & 7.545 & 14.265 \\[0.2cm]
    Transition $j \leftarrow j' $ \\[0.2cm]
    \hline\\[-0.25cm]
        27$\leftarrow$28   &     110.1   &   7.258    &     1.126      &        0.559  \\
        28$\leftarrow$29   &     122.4   &   8.068    &     1.252      &        0.622  \\
        29$\leftarrow$30   &     135.6   &   8.937    &     1.386      &        0.689  \\
        30$\leftarrow$31   &     149.7   &   9.866    &     1.531      &        0.761  \\
        31$\leftarrow$32   &     164.7   &   10.86   &     1.685       &        0.837  \\
        32$\leftarrow$33   &     180.7   &   11.91   &     1.848       &        0.918  \\
        33$\leftarrow$34   &     197.8   &   13.04   &     2.022       &        1.004 \\
        34$\leftarrow$35   &     215.8   &   14.22   &     2.207       &        1.096 \\
        35$\leftarrow$36   &     235.0   &   15.48   &     2.403       &        1.193 \\
        36$\leftarrow$37   &     255.2   &   16.82   &     2.610       &        1.296 \\
        37$\leftarrow$38   &     276.5   &   18.23   &     2.828       &        1.405 \\
        38$\leftarrow$39   &     299.0   &   19.71   &     3.058       &        1.519 \\
        39$\leftarrow$40   &     322.8   &   21.27   &     3.301       &        1.639 \\
        40$\leftarrow$41   &     347.7   &   22.91   &     3.556       &        1.766\\[0.15cm]
        \hline\hline
    \end{tabular}
\end{table}

From the data reported in  Table \ref{tab:Ecoeffs}, we can make the following observations:

(i) progressing along the series from the shorter to the longer anionic term, the value of $B_e$ becomes increasingly smaller and is reduced by more than one order of magnitude in going from  C$_3$N$^-$ to C$_{10}$H$^-$;

(ii) the corresponding permanent dipole moment, on the other hand, increases markedly from the shortest to the longest term in the series, varying in size by a factor of five;

(iii) since these two parameters appear with different exponents in equation (\ref{eq:dip_tran_rot}) it turns out that the size of the rotational constants dominate the values of the coefficients. Hence, they are seen in Table \ref{tab:Ecoeffs} to decrease by two orders of magnitude from the smallest to the longest linear chain in that table.

 To better understand the role of such variations in the regions where these anions are observed, it is useful to recall  another important indicator for describing the rotational state populations of the anions observed in Molecular Clouds: the LTE (Local Thermodynamic Equilibrium) assumption, a global condition which varies depending on the relative sizes of the local particle densities in the environments of interest and on the rotational level initial state distributions selected as the most likely in that environment. 

The features of the above indicator can be embodied in the analysis of the corresponding critical density values as recently shown, for example,  in \citep{LGS2023}. The definition of a critical density is given as:

\begin{equation}
n_{\text{crit}}^i(T) = \frac{A_{ij}}{\sum_{j\neq i} k_{ij}(T) }
\label{eq.critD}
\end{equation}

where the $n_{\text{crit}}^i$(T)  for any ith rotational level  is obtained by giving equal weight to the effects of the collision-induced  and the spontaneous emission processes, both quantities discussed and computed in previous Sections of this work. It is also worth mentioning  that there is no sum  over $ j$ values for the numerator in  equation (\ref{eq.critD}) since the
dipole-driven contribution is by far the largest from any
 given initial  state, so that values with  $\Delta j$ larger than 1 are essentially negligible .
 
Although it is well known that  particle densities  vary widely in different ISM environments we also know that several general models of the conditions in the molecular clouds suggest ranges going from the situation of the diffuse clouds (around 10$^4$ cm$^{-3}$) to dense clouds (around 10$^{10}$$-$10$^{12}$ cm$^{-3}$) as discussed, e.g.,  in \citep{06GoBoGi,Agundez2010} and in the references reported there. More specifically, the  conditions in the molecular clouds where the present anions have been detected indicate their densities expected to be around 10$^4$$-$10$^5$ cm$^{-3}$. Hence, from the de-excitation rates of Section \ref{sec:3.1} for C$_7$N$^-$, found to be around  10$^{-9}$ cm$^3$ molecule$^{-1}$ s$^{-1}$,  the  expected particle densities  generate a collisional de-excitation efficiency, in units of s$^{-1}$, of about 10$^{-4}$.
 
The $n_{\text{crit}}^i$(T) values from  equation (\ref{eq.critD}) are, in the case of the C$_7$N$^-$ anion, around 10$^{3}$$-$10$^{4}$ cm$^{-3}$. This size for $n_{\text{crit}}^i$(T) is clearly very close to the expected particle densities in the Molecular Clouds  where observations were made, thereby indicating that the rotational populations of that anion should be close to LTE conditions thanks to the large values of the collision-induced de-excitation rate coefficients. This is a useful suggestion since, to correctly interpret their detection, a complex analysis is necessary, and has to take into account as much as possible the validity of a non-local thermodynamic equilibrium (non-LTE) conditions of the emitting media (e.g. when energy level populations deviate from a Boltzman distribution). To confirm such assumptions  requires proper state-to-state collisional data for the excitation and de-excitation processes of the molecular levels, a task which we have tackled in the present study.

\section{Summary and conclusions}

In the present work,  we have reported for the first time  accurate ab initio PESs for the C$_7$N$^-$ and the C$_{10}$H$^-$ anions interacting with H$_2$. For both systems, the PESs were computed at the CCSD(T)-F12b/aug-cc-pVTZ level of theory, corrected for Basis Set Superposition Error and extrapolated into the long-range region by using the analytic expressions of the leading interaction terms (see earlier Sections). The validity of using the CCSD(T)-F12b procedure for these ionic interactions has been discussed many times in the literature; and therefore is not repeated here. Suffice it to say that a series of comparison with the same calculations at the CBS limit showed differences mostly around the attractive well regions and only of about 1-2$\%$. 

An accurate ANN fit of the orientation-averaged, 2D PES for the H$_2$ partner was carried out and also reported in detail in the earlier Section for the case of the  C$_7$N$^-$$-$H$_2$ interaction, while for the C$_{10}$H$^-$$-$H$_2$ system the radial coefficients of the 2D multipolar expansions were generated by initial numerical quadrature over the raw points and then extended into the long-range region by numerical extrapolation using the same analytic expression used for the other system. Both interactions were then employed using their multipolar expansion coefficients, representing their anisotropic interactions, for the 2D reduced-dimensionality representation of theis PESs (see earlier Sections). 

Because of the large number of rotational states which are actually coupled in these target species during the quantum dynamics, only exact CC  calculations were  carried out for the smaller C$_7$N$^-$ anion while, for the longer C$_{10}$H$^-$ anion, we resorted to using  the full CC coupling scheme for  the lower energies ( up to 150cm$^{-1}$) and the HD dynamical approximation  as the collision energies were increased, as described in detail in the previous Sections. The latter quantities were then used to obtain state-to-state inelastic rate coefficients as a function of $T$.  To reliably extend the range of temperatures for the computed rate coefficients up to 50 K, we have calculated the cross sections, for all transitions, up to 400 cm$^{-1}$. The numerical quadrature hence yielded  converged rate coefficient values within a few percentages. The transitions considered for the state-changing dynamics are the same as detected in the ISM for the C$_7$N$^-$ anion and discussed in \citep{Cern2023}. We have used the same type of transitions for the longer anion (C$_{10}$H$^-$) to carry out a more significant comparison of their behavior.

The present calculations have confirmed that the  state-to-state rate coefficients for C$_7$N$^-$$-$H$_2$  are  larger than those for the C$_7$N$^-$$-$He over the whole range of transitions examined, a result often reported in the earlier literature for such chains but seldom actually checked with exact calculations. 
We had already found in our earlier work \citep{BGG23}  that the  integral cross sections and rate coefficients for rotational transitions in C$_5$N$^-$ were consistently larger than the earlier results  for  C$_3$N$^-$ (e.g. see: \citep{Lara-Moreno2017, Lara-Moreno2019}. This is directly attributed to the larger anisotropy of the interacting PESs for C$_5$N$^-$ and to its dramatically  smaller rotational constant, when compared to the same features  for C$_3$N$^-$, as we have discussed in detail in the previous Section. The extension of such a comparison to the present calculations involving longer linear chains and considering the two largest polyynes observed so far in the ISM, have clearly shown that the rotation state-changing  probabilities for the longest anionic polyyne so far detected are indeed the largest ones in comparison with all the other terms of the series and indicate a very efficient  energy transfer mechanism acting within the collision-driven state-changing processes involving these systems, especially the longest chain anion: C$_{10}$H$^-$.

In the recent observations on the C$_{10}$H$^-$ rotation emission lines discussed in Ref. \citep{Cern2023}, a dense forest of rotational lines was  observed, thereby suggesting that the anion was present in that cold dark core in a variety of rotational excited states. The collision-driven population changes of those excited states  becomes a realistic evolutionary path  when we consider the rather large values of the rotational state-changing rates found in the present work from collisions  with H$_2$. Since such a partner is considered to be the most abundant neutral species both in CSEs and in molecular clouds, we can therefore argue that even the detection of the title anion  in cold molecular cores could be made  possible by the existence of these large collision-driven state-changing paths  found from the calculations  of the present study. Such findings also indicate  that these molecular anions are  likely to be close to a LTE  situation as far as their rotational state populations are concerned, although more specific kinetics modeling studies are still needed to confirm this suggestion. In any event, the collision paths to forming rotational excited states are found by the present calculations to be very effective energy-transfer mechanisms and therefore to be important reactions for modeling energy distributions within chemical kinetics studies involving the linear anionic polyynes.

\section*{Acknowledgements}

L.G.-S. and A.M.S.D. gratefully acknowledge grants PID2020-113147GA-I00 and PID2021-122839NB-I00 funded by Spanish Ministry of Science and Innovation
(MCIN/AEI/10.13039/MCIN/AEI/10.13039/501100011033). 
S.R. and U.L. thank NISER Bhubaneswar for providing computational facilities. 
F.A.G. acknowledges the support of the Computing Center of the Innsbruck University, where some of the present calculations were carried out. 
This research has also made use of the high performance computing resources of the Castilla y León Supercomputing Center (SCAYLE, \hyperlink{http://www.scayle.es}{www.scayle.es}), financed by the European Regional Development Fund (ERDF). N. S. thanks the Indian National Science Academy, New Delhi for the award of INSA Distinguished Professorship.

\section*{Data Availability}

The Supplementary Information folder contains the following material on our original data, made available to the readers of this paper:

\begin{itemize}

\item[-] Potential energy surface for the (C$_7$N$^-$/H$_2$) system provided as expansion multipolar coefficients as a function of ($R$, $\theta$); 
\item[-] Potential energy surface for the (C$_{10}$H$^-$/H$_2$) system provided as expansion multipolar coefficients as a function of ($R$, $\theta$); 
\item[-] The original ab initio points related to the original grids from which the two 2D PESs for both systems discussed in the present work have been obtained;

\item[-] The FORTRAN subroutine for the C$_7$N$^-$ $-$ H$_2$ interaction potential obtained by the best  ANN fit of the ab initio PES.  

\item[-] list of rotationally inelastic cross section and rate coefficient values for the (C$_7$N$^-$-H$_2$) system;

\item[-] list of rotationally inelastic cross section and rate coefficient values for the (C$_{10}$H$^-$/H$_2$) system;

\end{itemize}



\bibliographystyle{mnras}
\bibliography{mnras_draft.bib}

\begin{thebibliography}{}
\makeatletter
\relax
\def\mn@urlcharsother{\let\do\@makeother \do\$\do\&\do\#\do\^\do\_\do\%\do\~}
\def\mn@doi{\begingroup\mn@urlcharsother \@ifnextchar [ {\mn@doi@} {\mn@doi@[]}}
\def\mn@doi@[#1]#2{\def\@tempa{#1}\ifx\@tempa\@empty \href {http://dx.doi.org/#2} {doi:#2}\else \href {http://dx.doi.org/#2} {#1}\fi \endgroup}
\def\mn@eprint#1#2{\mn@eprint@#1:#2::\@nil}
\def\mn@eprint@arXiv#1{\href {http://arxiv.org/abs/#1} {{\tt arXiv:#1}}}
\def\mn@eprint@dblp#1{\href {http://dblp.uni-trier.de/rec/bibtex/#1.xml} {dblp:#1}}
\def\mn@eprint@#1:#2:#3:#4\@nil{\def\@tempa {#1}\def\@tempb {#2}\def\@tempc {#3}\ifx \@tempc \@empty \let \@tempc \@tempb \let \@tempb \@tempa \fi \ifx \@tempb \@empty \def\@tempb {arXiv}\fi \@ifundefined {mn@eprint@\@tempb}{\@tempb:\@tempc}{\expandafter \expandafter \csname mn@eprint@\@tempb\endcsname \expandafter{\@tempc}}}

\bibitem[\protect\citeauthoryear{{Ag\'undez, M.}, {Cernicharo, J.}, {Gu\'elin, M.}, {Gerin, M.}, {McCarthy, M. C.}  \& {Thaddeus, P.}}{{Ag\'undez, M.} et~al.}{2008}]{Agundez2008}
{Ag\'undez, M.} {Cernicharo, J.} {Gu\'elin, M.} {Gerin, M.} {McCarthy, M. C.}  {Thaddeus, P.} 2008, \mn@doi [A\&A] {10.1051/0004-6361:20078985}, 478, L19

\bibitem[\protect\citeauthoryear{{Ag\'undez, M.} et~al.,}{{Ag\'undez, M.} et~al.}{2010}]{Agundez2010}
{Ag\'undez, M.} et~al., 2010, \mn@doi [A\&A] {10.1051/0004-6361/201015186}, 517, L2

\bibitem[\protect\citeauthoryear{Arthurs \& Dalgarno}{Arthurs \& Dalgarno}{1960}]{60ArDaxx}
Arthurs A.~M.,  Dalgarno A.,  1960, \mn@doi [Proc. R. Soc. A] {10.1098/rspa.1960.0125}, 256, 540

\bibitem[\protect\citeauthoryear{Biswas, Rashmi  \& Lourderaj}{Biswas et~al.}{2020}]{Biswas}
Biswas R.,  Rashmi R.,   Lourderaj U.,  2020, Resonance, 25, 59

\bibitem[\protect\citeauthoryear{Biswas et~al.,}{Biswas et~al.}{2023}]{BGG23}
Biswas R.,  et~al., 2023, \mn@doi [MNRAS] {https://doi.org/10.1093/mnras/stad1261}, 522, 5775

\bibitem[\protect\citeauthoryear{Botschwina \& Oswald}{Botschwina \& Oswald}{2008}]{Botschwina}
Botschwina P.,  Oswald R.,  2008, \mn@doi [J. Chem. Phys.] {https://doi.org/10.1063/1.2949093}, 129, 044305

\bibitem[\protect\citeauthoryear{Brown \& Carrington}{Brown \& Carrington}{2003}]{03BrCaxx}
Brown J.~M.,  Carrington A.,  2003, Rotational Spectroscopy of Diatomic Molecules.
Cambridge University Press, Cambridge

\bibitem[\protect\citeauthoryear{Br{\"u}nken, Gupta, Gottlieb, McCarthy  \& Thaddeus}{Br{\"u}nken et~al.}{2007}]{Brunken2007}
Br{\"u}nken S.,  Gupta H.,  Gottlieb C.,  McCarthy M.,   Thaddeus P.,  2007, \mn@doi [ApJ] {10.1086/520703}, 664, L43

\bibitem[\protect\citeauthoryear{Buonomo, Gianturco, de Lara-Castells, Delgado-Barrio, Miret-Artés  \& Villarreal}{Buonomo et~al.}{1997}]{Buonomo}
Buonomo E.,  Gianturco F.~A.,  de Lara-Castells M.~P.,  Delgado-Barrio G.,  Miret-Artés S.,   Villarreal P.,  1997, \mn@doi [J. Chem. Phys.] {10.1063/1.473976}, 106, 1718

\bibitem[\protect\citeauthoryear{{Cernicharo, J.}, {Gu\'elin, M.}, {Ag\'undez, M.}, {Kawaguchi, K.}, {McCarthy, M.}  \& {Thaddeus, P.}}{{Cernicharo, J.} et~al.}{2007}]{Cernicharo2007}
{Cernicharo, J.} {Gu\'elin, M.} {Ag\'undez, M.} {Kawaguchi, K.} {McCarthy, M.}  {Thaddeus, P.} 2007, \mn@doi [A\&A] {10.1051/0004-6361:20077415}, 467, L37

\bibitem[\protect\citeauthoryear{{Cernicharo, J.}, {Marcelino, N.}, {Pardo, J. R.}, {Ag\'undez, M.}, {Tercero, B.}, {de Vicente, P.}, {Cabezas, C.}  \& {Berm\'udez, C.}}{{Cernicharo, J.} et~al.}{2020}]{Cernicharo2020}
{Cernicharo, J.} {Marcelino, N.} {Pardo, J. R.} {Ag\'undez, M.} {Tercero, B.} {de Vicente, P.} {Cabezas, C.}  {Berm\'udez, C.} 2020, \mn@doi [A\&A] {10.1051/0004-6361/202039231}, 641, L9

\bibitem[\protect\citeauthoryear{Cernicharo, Guélin, Agúndez, McCarthy  \& Thaddeus}{Cernicharo et~al.}{2008}]{Cernicharo2008}
Cernicharo J.,  Guélin M.,  Agúndez M.,  McCarthy M.~C.,   Thaddeus P.,  2008, \mn@doi [ApJ] {10.1086/593183}, 688, L83–L86

\bibitem[\protect\citeauthoryear{Cernicharo et~al.,}{Cernicharo et~al.}{2023}]{Cern2023}
Cernicharo J.,  et~al., 2023, \mn@doi [A\&A] {doi.org/10.1051/0004-6361/202245816}, 670, L19

\bibitem[\protect\citeauthoryear{Gayatri \& Sathyamurthy}{Gayatri \& Sathyamurthy}{1980}]{GS80}
Gayatri C.,  Sathyamurthy N.,  1980, Chem.Phys., 48, 227

\bibitem[\protect\citeauthoryear{Gianturco}{Gianturco}{1979}]{79FaGxx}
Gianturco F.,  1979, Lect.Notes Chem., Springer Verlag, Berlin

\bibitem[\protect\citeauthoryear{Giri, González-Sánchez, Biswas, Yurtsever, Gianturco, Sathyamurthy, Lourderaj  \& Wester}{Giri et~al.}{2022}]{Giri}
Giri K.,  González-Sánchez L.,  Biswas R.,  Yurtsever E.,  Gianturco F.~A.,  Sathyamurthy N.,  Lourderaj U.,   Wester R.,  2022, \mn@doi [J. Phys. Chem. A] {10.1021/acs.jpca.1c10309}, 126, 2244

\bibitem[\protect\citeauthoryear{Gonz\'alez-S\'anchez, Bodo  \& Gianturco}{Gonz\'alez-S\'anchez et~al.}{2006}]{06GoBoGi}
Gonz\'alez-S\'anchez L.,  Bodo E.,   Gianturco F.~A.,  2006, \mn@doi [Phys. Rev. A] {10.1103/PhysRevA.73.022703}, 73, 022703

\bibitem[\protect\citeauthoryear{{Gonz\'{a}lez-S\'{a}nchez}, {Gianturco}, {Carelli}  \& {Wester}}{{Gonz\'{a}lez-S\'{a}nchez} et~al.}{2015}]{15GoGiCa}
{Gonz\'{a}lez-S\'{a}nchez} L.,  {Gianturco} F.~A.,  {Carelli} F.,   {Wester} R.,  2015, \mn@doi [New. J. Phys.] {10.1088/1367-2630/17/12/123003}, 17, 123003

\bibitem[\protect\citeauthoryear{Gonz{\'{a}}lez-S{\'{a}}nchez, Mant, Wester  \& Gianturco}{Gonz{\'{a}}lez-S{\'{a}}nchez et~al.}{2020}]{GonzalezSanchezL}
Gonz{\'{a}}lez-S{\'{a}}nchez L.,  Mant B.~P.,  Wester R.,   Gianturco F.~A.,  2020, \mn@doi [ApJ] {10.3847/1538-4357/ab94a0}, 897, 75

\bibitem[\protect\citeauthoryear{González-Sánchez et~al.,}{González-Sánchez et~al.}{2024}]{LGS2023}
González-Sánchez L.,  et~al., 2024, \mn@doi [ApJ] {doi.org/10.3847/1538-4357/ad055e}, 960, 40

\bibitem[\protect\citeauthoryear{Green}{Green}{1975}]{Green}
Green S.,  1975, \mn@doi [J. Chem. Phys.] {10.1063/1.430752}, 62, 2271

\bibitem[\protect\citeauthoryear{Hutson}{Hutson}{1994}]{94JmHxx}
Hutson J.,  1994, \mn@doi [Comp. Phys. Comm.] {https://doi.org/10.1016/0010-4655(94)90200-3}, 84, 1

\bibitem[\protect\citeauthoryear{Hutson \& Sueur}{Hutson \& Sueur}{2019a}]{MOLSCAT2}
Hutson J.~M.,  Sueur C. R.~L.,  2019a, MOLSCAT a program for non-reactive quantum scattering calculation on atomic and molecular collisions, \url {https://github.com/molscat/molscat}

\bibitem[\protect\citeauthoryear{Hutson \& Sueur}{Hutson \& Sueur}{2019b}]{MOLSCAT}
Hutson J.~M.,  Sueur C. R.~L.,  2019b, \mn@doi [Comput. Phys. Commun.] {10.1016/j.cpc.2019.02.014}, 241, 9

\bibitem[\protect\citeauthoryear{Jerosimi\'{c}, Gianturco  \& Wester}{Jerosimi\'{c} et~al.}{2018}]{18JeGiWe.Cnm}
Jerosimi\'{c} S.~V.,  Gianturco F.~A.,   Wester R.,  2018, \mn@doi [Phys. Chem. Chem. Phys.] {10.1039/C7CP05573K}, 20, 5490

\bibitem[\protect\citeauthoryear{K{\l}os \& Lique}{K{\l}os \& Lique}{2011}]{Klos2011}
K{\l}os J.,  Lique F.,  2011, \mn@doi [MNRAS] {https://doi.org/10.1111/j.1365-2966.2011.19481.x}, 418, 271

\bibitem[\protect\citeauthoryear{Kolos \& Wolniewicz}{Kolos \& Wolniewicz}{1967}]{kol1967}
Kolos W.,  Wolniewicz L.,  1967, \mn@doi [J. Chem. Phys.] {10.1063/1.1840870}, 46, 1426

\bibitem[\protect\citeauthoryear{Kouri}{Kouri}{1975}]{kouri1975decoupling}
Kouri D.~J.,  1975, \mn@doi [Chem. Phys. Lett.] {10.1016/0009-2614(75)85095-0}, 31, 599

\bibitem[\protect\citeauthoryear{Kouri \& Hoffman}{Kouri \& Hoffman}{1997}]{97KoHoffxx}
Kouri D.,  Hoffman D.,  1997, \mn@doi [Truhlar D.G., Simon B. (eds) Multiparticle Quantum Scattering With Applications to Nuclear, Atomic and Molecular Physics] {https://doi.org/10.1007/978-1-4612-1870}, 89, Springer, New York, NY

\bibitem[\protect\citeauthoryear{Lara-Moreno, Stoecklin  \& Halvick}{Lara-Moreno et~al.}{2017}]{Lara-Moreno2017}
Lara-Moreno M.,  Stoecklin T.,   Halvick P.,  2017, \mn@doi [MNRAS] {https://doi.org/10.1093/mnras/stx434}, 467, 4174

\bibitem[\protect\citeauthoryear{Lara-Moreno, Stoecklin  \& Halvick}{Lara-Moreno et~al.}{2019}]{Lara-Moreno2019}
Lara-Moreno M.,  Stoecklin T.,   Halvick P.,  2019, \mn@doi [MNRAS] {https://doi.org/10.1093/mnras/stz860}, 486, 414

\bibitem[\protect\citeauthoryear{L\'opez-Dur\'an, Bodo  \& Gianturco}{L\'opez-Dur\'an et~al.}{2008}]{08LoBoGi}
L\'opez-Dur\'an D.,  Bodo E.,   Gianturco F.~A.,  2008, \mn@doi [Comput. Phys. Commun.] {10.1016/j.cpc.2008.07.017}, 179, 821

\bibitem[\protect\citeauthoryear{Mant, Franz, Wester  \& Gianturco}{Mant et~al.}{2021}]{BM2021}
Mant B.,  Franz J.,  Wester R.,   Gianturco F.~A.,  2021, \mn@doi [Mol. Phys.] {10.1080/00268976.2021.1938267}, 119, 1

\bibitem[\protect\citeauthoryear{Martinazzo, Bodo  \& Gianturco}{Martinazzo et~al.}{2003}]{03MaBoGi}
Martinazzo R.,  Bodo E.,   Gianturco F.~A.,  2003, \mn@doi [Comput. Phys. Commun.] {10.1016/S0010-4655(02)00737-3}, 151, 187

\bibitem[\protect\citeauthoryear{McCarthy, Gottlieb, Gupta  \& Thaddeus}{McCarthy et~al.}{2006}]{McCarthy2006}
McCarthy M.~C.,  Gottlieb C.~A.,  Gupta H.,   Thaddeus P.,  2006, \mn@doi [ApJ] {10.1086/510238}, 652, L141

\bibitem[\protect\citeauthoryear{Polanyi \& Woodall}{Polanyi \& Woodall}{1972}]{PW72}
Polanyi J.,  Woodall K.,  1972, J.Chem.Phys., 56, 1563

\bibitem[\protect\citeauthoryear{Raff, Komanduri, Hagan  \& Bukkapatnam}{Raff et~al.}{2012}]{raff2012}
Raff L.~M.,  Komanduri R.,  Hagan M.,   Bukkapatnam S. T.~S.,  2012, Neural Networks in Chemical Reaction Dynamics.
Oxford Univ. Press, Oxford

\bibitem[\protect\citeauthoryear{Remijan, Hollis, Lovas, Cordiner, Millar, Markwick-Kemper  \& Jewell}{Remijan et~al.}{2007}]{Remijan2007}
Remijan A.~J.,  Hollis J.~M.,  Lovas F.~J.,  Cordiner M.~A.,  Millar T.~J.,  Markwick-Kemper A.~J.,   Jewell P.~R.,  2007, \mn@doi [ApJ] {https://doi.org/10.1086/520704}, 664, L47

\bibitem[\protect\citeauthoryear{Remijan et~al.,}{Remijan et~al.}{2023}]{Rem2023}
Remijan A.,  et~al., 2023, \mn@doi [ApJL] {doi.org/10.3847/2041-8213/acb648}, 944, L45

\bibitem[\protect\citeauthoryear{Sarkar \& Bhattacharyya}{Sarkar \& Bhattacharyya}{2017}]{Sarkar}
Sarkar K.,  Bhattacharyya S.~P.,  2017, Soft-computing in Physical and Chemical Sciences: A Shift in Computing Paradigm.
CRC Press

\bibitem[\protect\citeauthoryear{Satta, Gianturco, Carelli  \& Wester}{Satta et~al.}{2015}]{15SaGiCa.LM}
Satta M.,  Gianturco F.~A.,  Carelli F.,   Wester R.,  2015, \mn@doi [ApJ] {10.1088/0004-637X/799/2/228}, 799, 228

\bibitem[\protect\citeauthoryear{Secrest}{Secrest}{1979}]{79Secrxx}
Secrest D.,  1979, \mn@doi [Bernstein R.B. (eds) Atom - Molecule Collision Theory] {https://doi.org/10.1007/978-1-4613-2913-8}, Plenum, New York

\bibitem[\protect\citeauthoryear{{Taylor}}{{Taylor}}{2006}]{Taylor2006}
{Taylor} J.~R.,  2006, {Scattering Theory The Quantum Theory of Nonrelativistic Collisions}.
Dover

\bibitem[\protect\citeauthoryear{{Thaddeus}, {Gottlieb}, {Gupta}, {Br{\"u}nken}, {McCarthy}, {Ag{\'u}ndez}, {Gu{\'e}lin}  \& {Cernicharo}}{{Thaddeus} et~al.}{2008}]{Thaddeus2008}
{Thaddeus} P.,  {Gottlieb} C.~A.,  {Gupta} H.,  {Br{\"u}nken} S.,  {McCarthy} M.~C.,  {Ag{\'u}ndez} M.,  {Gu{\'e}lin} M.,   {Cernicharo} J.,  2008, \mn@doi [\apj] {10.1086/528947}, 677, 1132

\bibitem[\protect\citeauthoryear{Werner, Knowles, Knizia, Manby  \& Sch\"utz}{Werner et~al.}{2012}]{molpro}
Werner H.-J.,  Knowles P.~J.,  Knizia G.,  Manby F.~R.,   Sch\"utz M.,  2012, \mn@doi [WIREs Comput. Mol. Sci.] {10.1002/wcms.82}, 2, 242

\makeatother
\end{thebibliography}



%
%
%

\bsp	
\label{lastpage}
\end{document}